# Tellurium trioxide as an n-type wide-gap semiconductor with ultrahigh work functions from first principles

Tianwei Wang [a,b,*], Teruya Nagafuji [b], Kota Hanzawa [b], Hidenori Hiramatsu [b,c], and Fumiyasu Oba [a,b,c,*]

[a.] Kanagawa Institute of Industrial Science and Technology (KISTEC), Ebina 243-0435, Japan

[b.] Materials and Structures Laboratory, Institute of Integrated Research, Institute of Science Tokyo, Yokohama 226-8501, Japan

[c.] MDX Research Center for Element Strategy, Institute of Integrated Research, Institute of Science Tokyo, Yokohama 226-8501, Japan

[*] Corresponding authors.

E-mail addresses: wang.t@msl.iir.isct.ac.jp (T. Wang), oba.f@msl.iir.isct.ac.jp (F. Oba)

## ABSTRACT

The exploration of next-generation wide-gap semiconductors with extreme work functions is highly desirable for the continued advancement of electronic and optoelectronic devices. By using first principles calculations, we show that $\beta$-$TeO_3$, with a wide band gap exceeding 3 eV, emerges as a promising n-type semiconductor exhibiting ultrahigh work functions of up to ~8 eV. Intentionally or unintentionally incorporated interstitial H, Li, Na, Cu, Mg, and Zn impurities and substitutional F impurities are identified as shallow donors, whereas O vacancies act as minor carrier-compensating defects that locally trap electrons at neighbouring Te sites. The work function is evaluated across a series of low-index nonpolar surfaces. The presence of ordered O vacancies is energetically favoured on these surfaces and tends to reduce the corresponding work functions. These findings underscore the significance of precise control over the O chemical potential to suppress surface O-vacancy formation, thereby preserving the exceptional electronic performance of $\beta$-$TeO_3$ associated with ultrahigh work functions. Collectively, the results highlight the potential of $\beta$-$TeO_3$ for hole injection/extraction/transport layers as well as transparent electrodes for applications such as light-emitting devices and photovoltaic cells, while providing useful guidance on carrier doping and surface tuning for these practical applications through elucidating the effects of the point-defect and surface configurations on the electronic properties.

## 1. Introduction

Wide-gap semiconductors with high work functions exceeding those of prototypical electrode materials such as elemental Pt (5.65 eV) [1] and Sn-doped $In_2O_3$ (about 4.7 eV for as-prepared surfaces and 5.6 eV when plasma-treated) [2,3] are of significant interest for applications such as hole injection/extraction/transport layers and electrodes in various devices such as light-emitting diodes and photovoltaic cells. Among these, $MoO_x$ has been extensively studied as an n-type semiconductor [4–9]: notably high work functions reported in the range of 6.8–6.9 eV for $\alpha$-$MoO_3$ with the conduction band minima (CBM) located 6.6–6.7 eV below the vacuum level, corresponding to their electron affinity (EA) values [8–10], whereas lower work functions down to 4.7 eV for $MoO_x$ ($x < 3$) [11,12]. The high work function of $MoO_x$ ($x \approx 3$) has been theoretically attributed to the $d^0$ electronic structure of Mo(VI), together with the substantial dipole contributions arising from surface Mo-O bonds [13]. A systematic first-principles investigation into binary and ternary oxide surfaces indicates that a high cation oxidation state, +6 in the case of $MoO_3$ with an ideal stoichiometry, is favourable to lower the valence band maximum (VBM) and the CBM with respect to the vacuum level, namely, to increase the ionisation potential (IP) and the EA [14–16]. Consequently, high work functions are expected for such oxides. Motivated by these benchmarking characteristics of $MoO_x$ and insights into the chemical tendencies governing the band alignment of oxides, we propose $\beta$-$TeO_3$ as a promising n-type wide-gap semiconductor with ultrahigh work functions.

$\beta$-$TeO_3$ adopts the trigonal $R\bar{3}c$ space group (No. 167) [17]. The synthesis of $\beta$-$TeO_3$ has been reported mainly by heating the telluric acid $H_6TeO_6$ [18–22], typically within a narrow temperature range of 350–400 °C, while amorphous $TeO_3$ and $\alpha$-$TeO_2$ (or $Te_2O_5$ depending on the temperature) phases have been found at lower and higher temperatures, respectively. $\beta$-$TeO_3$ has also been obtained by heating amorphous $TeO_3$ in a sealed environment, but its decomposition into $Te_2O_5$ was observed at 450 °C, which would further decompose into $\alpha$-$TeO_2$ at a higher temperature [18,19]. Besides, multiple equilibrium phase diagram analyses [23,24] that take the above into consideration indicate $\beta$-$TeO_3$ as a thermodynamically stable phase.

The crystal structure of $\beta$-$TeO_3$ is provided in Fig. 1a. Each Te cation in the +6 oxidation state is octahedrally coordinated by six equivalent O anions with an O-Te-O bond angle of 89.1°, forming corner-sharing $TeO_6$ octahedra. Each bridging O anion is bonded to two equivalent Te cations, giving rise to a Te-O-Te angle of 131.3°, yielding lattice parameters of $a_p$ = 5.184 Å and $\alpha$ = 56.42° for the primitive/rhombohedral cell, and $a$ = 4.901 Å and $c$ = 13.030 Å for the conventional/hexagonal cell [17]. Although the thin films of amorphous $TeO_3$ that demonstrate electrical conduction have been reported [25], followed by a more recent application of amorphous $TeO_3$ in p-channel field-effect transistors [26], insights into the fundamental and defect properties of crystalline $TeO_3$ remain lacking.

Herein, we present a systematic prediction of the bulk, defect, and surface properties of $\beta$-$TeO_3$ by using first principles calculations. Several impurities, including interstitial H, Li, Na, Cu, Mg, and Zn, as well as F substituting for O sites, emerge as shallow donors. Native O vacancies trap carrier electrons, but their formation can be energetically suppressed. The work functions are as high as ~8.0 eV for ideal $\beta$-$TeO_3$ surfaces but are lowered by O-vacancy formation, requiring a high O chemical potential to maintain their ultrahigh work functions. These findings suggest that $\beta$-$TeO_3$ is a promising n-type wide-gap semiconductor for hole injection/extraction/transport layers and transparent electrodes necessitating high work functions.

## 2. Computational methods

The first principles calculations were carried out adopting the projector augmented-wave (PAW) method [27] within the generalised Kohn–Sham framework [28], as implemented in the VASP code [29–31]. The PAW dataset details are available in Table S1 of the Supplementary materials. The exchange-correlation term was approximated by adopting the Heyd–Scuseria–Ernzerhof (HSE) hybrid functional [32–34]. A Fock exchange mixing parameter of 0.29 was employed to reproduce the experimental band gap of $TeO_3$. A plane-wave energy cutoff of 520 eV was chosen for the optimisation of primitive cells, including that of $TeO_3$ and the competing phases, where all cell dimensions were allowed to relax, while a reduced cutoff of 400 eV was applied for all subsequent calculations, including those on the supercells constructed from the fully optimised $TeO_3$ primitive cell. A Γ-centred 7×7×7 $k$-point mesh was used for optimising the primitive cell of $TeO_3$; a similar $k$-point density of over 4.0 Å was adopted for the optimisation of slab-model supercells and a reduced density of over 2.5 Å for point-defect supercells. Ionic positions were optimised until all force residuals were below 0.005, 0.03, and 0.01 eV/Å for the optimisation of primitive cells, point-defect supercells, and slab-model supercells, respectively. All atomic structures were visualised by the VESTA software [35].

With respect to the self-consistent response to a finite electric field [36,37], both electronic and ionic contributions to the static dielectric tensor were determined, with the latter obtained by further introducing ionic displacements. Electronic contributions to the frequency-dependent dielectric response were calculated within the independent-particle picture [38], contributing to the imaginary part of the dielectric function, whose real part was further obtained by the Kramers-Kronig transformation. Enhanced Γ-centred $k$-point meshes of 14×14×14 were utilised to calculate these dielectric properties and the density of states (DOS) based on the optimised $TeO_3$ primitive cell; band-averaged effective mass tensors were derived from the latter by using the BoltzTraP2 package [39]. The results of first principles calculations were post-processed and visualised utilising both pymatgen and vise packages [40,41].

The donor binding energy $E_b$ was approximated by the hydrogenic model in the ground-state Rydberg form [42]

$$E_{\mathrm{b}} = \left(\frac{m_{\mathrm{e}}^*}{m_0}\right)\frac{1}{\varepsilon_{\mathrm{r}}^2}\,\mathrm{Ry}, \tag{1}$$

where $m_{\mathrm{e}}^*$ is the spherical average of the calculated electron effective mass tensor and $\varepsilon_{\mathrm{r}}$ is the spherical average of the calculated static dielectric tensor relative to vacuum permittivity. Ry is the Rydberg energy of about 13.6 eV.

The point defects in $TeO_3$ were investigated using supercells constructed from a 2×2×1 expansion of the conventional cell. Point defects were introduced with their adjacent atoms randomly displaced by up to 0.2 Å from bulk coordinates to break the cell symmetry. The ShakeNBreak package [43,44] was additionally utilised to generate candidate atomic configurations for self-trapped electrons (STEs). Candidate interstitial positions were selected from the local minima of the electrostatic potential, those of the charge density, and the local maxima of the electron localisation function [45]. Cell dimensions were fixed for point-defect calculations, while all internal coordinates were optimised. Spin polarisation was considered in all point-defect calculations. The finite-size errors in electrostatics and their impact on defect formation energies were corrected by the extended Freysoldt–Neugebauer–Van de Walle (eFNV) scheme [46,47], which incorporates the anisotropic electrostatic screening effect derived from the calculated static dielectric tensor containing both electronic and ionic contributions. The pydefect package was employed to set up and interpret point-defect calculations [41].

The formation energy of a point defect $D$ in the charge state $q$ (denoted as $D^q$) was calculated by

$$E_{\mathrm{f}}[D^q] = E_{\mathrm{tot}}[D^q] + E_{\mathrm{eFNV}}[D^q] - E_{\mathrm{tot}}^{\mathrm{bulk}} - \textstyle\sum_i \Delta n_i\, \mu_i + q(\varepsilon_{\mathrm{VBM}} + \Delta\varepsilon_{\mathrm{F}}), \tag{2}$$

where $E_{\mathrm{tot}}[D^q]$ is the total energy of the defect-containing supercell and $E_{\mathrm{tot}}^{\mathrm{bulk}}$ is that of the corresponding bulk (perfect crystal) supercell. The change in the number of atoms of species $i$ due to introducing the point defect $D$ is represented by $\Delta n_i$ with the atomic chemical potential $\mu_i$. The energy level at the VBM is given by $\varepsilon_{\mathrm{VBM}}$ and the Fermi level $\Delta\varepsilon_{\mathrm{F}}$ is in reference to $\varepsilon_{\mathrm{VBM}}$. The term $E_{\mathrm{eFNV}}[D^q]$ represents the finite-size correction under the eFNV scheme. The chemical potentials $\mu_{\mathrm{Te}}$ and $\mu_{\mathrm{O}}$ in equilibrium are constrained by

$$\mu_{\mathrm{Te}} + 3\mu_{\mathrm{O}} = E_{\mathrm{TeO_3}}^{\mathrm{bulk}} \tag{3}$$

in which $E_{\mathrm{TeO_3}}^{\mathrm{bulk}}$ is the total energy per formula unit of bulk $TeO_3$ (perfect crystal).

Potential competing phases used to determine the limits of atomic chemical potentials were selected from the Materials Project [48], by using its containing database of thermodynamic stability [49–51], and particularly the convex hull approach [52]; those with relative formation energies of below 0.005 eV/atom above the convex hull in the Materials Project were chosen and re-calculated using the HSE hybrid functional as described in an earlier paragraph. Potential competing phases that contain Cu(II) with various degrees of magnetic ordering were considered in addition to the above, as summarised in Table S2 of the Supplementary materials.

The defect formation energies were utilised to estimate the defect concentrations at a finite temperature $T$ following the Boltzmann distribution given by

$$C[D^q] = N_{\mathrm{site}}[D^q] \cdot N_{\mathrm{spin}}[D^q] \cdot \exp(-E_{\mathrm{f}}[D^q]/k_{\mathrm{B}}T), \tag{4}$$

in which $N_{\mathrm{site}}$ and $N_{\mathrm{spin}}$ represent the site and spin degeneracies of a defect configuration, respectively, and $k_{\mathrm{B}}$ is the Boltzmann's constant. $N_{\mathrm{site}}$ was calculated by the number of symmetry operations of the perfect crystal divided by the number of remaining symmetry operations after a point defect was introduced and relaxed. $N_{\mathrm{spin}}$ was set to 1 for non-spin-polarised defect configurations and 2 for spin-polarised cases. The equilibrium carrier concentration and the corresponding Fermi level were then obtained by self-consistently solving the equation of charge neutrality:

$$p - n + \sum q \cdot C[D^q] = 0, \tag{5}$$

where the electron concentration $n$ above the CBM (whose energy level denoted by $\varepsilon_{\mathrm{CBM}}$) and the hole concentration $p$ below the VBM ($\varepsilon_{\mathrm{VBM}}$) were calculated from the DOS, denoted as $\mathcal{D}(\varepsilon)$, by the following:

$$n = \int_{\varepsilon_{\mathrm{CBM}}}^{\infty} \mathcal{D}(\varepsilon) \frac{1}{\exp\left(\frac{\varepsilon-\varepsilon_{\mathrm{F}}}{k_{\mathrm{B}}T}\right)+1} \mathrm{d}\varepsilon, \tag{6}$$

$$p = \int_{-\infty}^{\varepsilon_{\mathrm{VBM}}} \mathcal{D}(\varepsilon) \frac{1}{\exp\left(\frac{\varepsilon_{\mathrm{F}}-\varepsilon}{k_{\mathrm{B}}T}\right)+1} \mathrm{d}\varepsilon. \tag{7}$$

The supercells containing slabs terminated by low-index nonpolar surfaces (with or without O vacancies) were constructed by truncating HSE-optimised bulk structures. Each slab was sliced to maintain a minimum thickness of 21 Å and separated from its periodic images by vacuum regions of at least 15 Å thick along the surface normal. To reduce computational cost, the internal coordinates of all constructed slab-model supercells and the corresponding primitive cell were optimised using the Perdew–Burke–Ernzerhof functional revised for solids (PBEsol) [53], with all cell dimensions held fixed. Subsequently, their corresponding total energies and electronic structures were obtained via single-shot calculations utilising the same hybrid-functional setup as detailed above. EAs and IPs were calculated by the following [54,55],

$$EA = (\varepsilon_{\mathrm{vac}} - \varepsilon_{\mathrm{ref}}) - \varepsilon_{\mathrm{CBM-ref}}, \tag{8}$$

$$IP = (\varepsilon_{\mathrm{vac}} - \varepsilon_{\mathrm{ref}}) - \varepsilon_{\mathrm{VBM-ref}}. \tag{9}$$

The electrostatic reference level $\varepsilon_{\mathrm{ref}}$ and the vacuum level $\varepsilon_{\mathrm{vac}}$ were obtained by evaluating the respective macroscopic averages of the stabilised electrostatic potentials of bulk-like and vacuum regions of a slab-model supercell [56,57], sharing a common electrostatic reference level with $\varepsilon_{\mathrm{CBM}}$ and $\varepsilon_{\mathrm{VBM}}$, namely $\varepsilon_{\mathrm{CBM-ref}}$ and $\varepsilon_{\mathrm{VBM-ref}}$ from the corresponding bulk primitive cell. Since the formation of O vacancies is energetically favourable, both stoichiometric surfaces and those containing O vacancies have been examined. Their relative stability was evaluated by comparing their surface energies, defined as

$$E_{\mathrm{surf}} = \frac{1}{2A}\left(E_{\mathrm{tot}}^{\mathrm{slab}} - n_{\mathrm{Te}}\mu_{\mathrm{Te}} - n_{\mathrm{O}}\mu_{\mathrm{O}}\right), \tag{10}$$

where $E_{\mathrm{tot}}^{\mathrm{slab}}$ is the total energy of the slab model containing $n_{\mathrm{Te}}$ number of Te atoms and $n_{\mathrm{O}}$ number of O atoms, respectively. The in-plane surface area of each slab surface is given by $A$. The O-rich and O-poor conditions equivalent to those for the point-defect calculations were considered as the limits of the Te and O chemical potentials.

## 3. Results and discussion

### *3.1 Bulk properties of β-$TeO_3$*

Our first principles calculations adopting the HSE hybrid functional [32–34] accurately reproduce the experimental crystal structure of $TeO_3$. The calculated lattice parameters are $a_{\mathrm{p}}$ = 5.181 Å and $\alpha$ = 56.94° for the primitive cell, and $a$ = 4.939 Å and $c$ = 12.977 Å for the conventional cell. The calculated band structure in Fig. 1b reveals an indirect band gap of 3.13 eV, while the corresponding experimental value is 3.1 eV as we have estimated by analysing the diffuse reflectance spectrum reported by Roginskii *et al.* [58] (see Fig. S1 in the Supplementary materials). The CBM is located at the Γ point and the VBM between the Γ and L points. The topmost valence band is nearly flat and exhibits a local minimum at the Γ point, with tiny increases along both Γ-L and Γ-S directions. The electronic states just below the VBM originate almost entirely from O-2p orbitals with negligible Te contributions. In contrast, the bottommost conduction band exhibits a relatively large dispersion, which is primarily composed of Te-5s orbitals with O-2p contributions. The calculated optical absorption spectrum is shown in Fig. 1c. A gradual increase in the absorption coefficient near a photon energy of ~3.1 eV is observed, reflecting the indirect-type band structure with a small difference between the minimum indirect and direct gaps. Therefore, $\beta$-$TeO_3$ is expected to be almost transparent in thin-film the form, which is advantageous for applications as transparent electrodes and carrier-transport layers.

The band-averaged effective masses of carrier electrons and holes are provided in Table 1. $TeO_3$ features relatively low electron effective masses of 0.64 and 0.44 for the directions normal (*xx*) and parallel (*zz*) to the *c* axis, respectively. These values are only a few times as large as those of representative oxide semiconductors, such as ZnO (0.15$m_0$ theoretically estimated and 0.24$m_0$ experimentally measured), $In_2O_3$ (0.17$m_0$ and 0.30$m_0$), and $SnO_2$ (0.17$m_0$ and 0.35$m_0$) [59–62], indicating the suitability of $TeO_3$ as an n-type semiconductor. In addition, the spherical average of the electron effective masses yields an approximate binding energy of 64 meV for a hydrogenic donor state as in Eq. (1). This is comparable to those in commonly used n-type semiconductors (46 meV in P-doped Si and 55 meV in Ga-doped ZnO) [63,64], indicating significant potential of shallow donor formation in $TeO_3$. Note that the exact binding energy of a hydrogenic donor state depends on the dopant type and its estimation by first principles calculations requires huge computational cells containing at least tens of thousands of atoms, depending on the spatial distribution of the donor states [65,66], which is beyond the scope of this study. On the other hand, high hole effective masses are found, making shallow hydrogenic acceptor state formation unlikely in $TeO_3$.

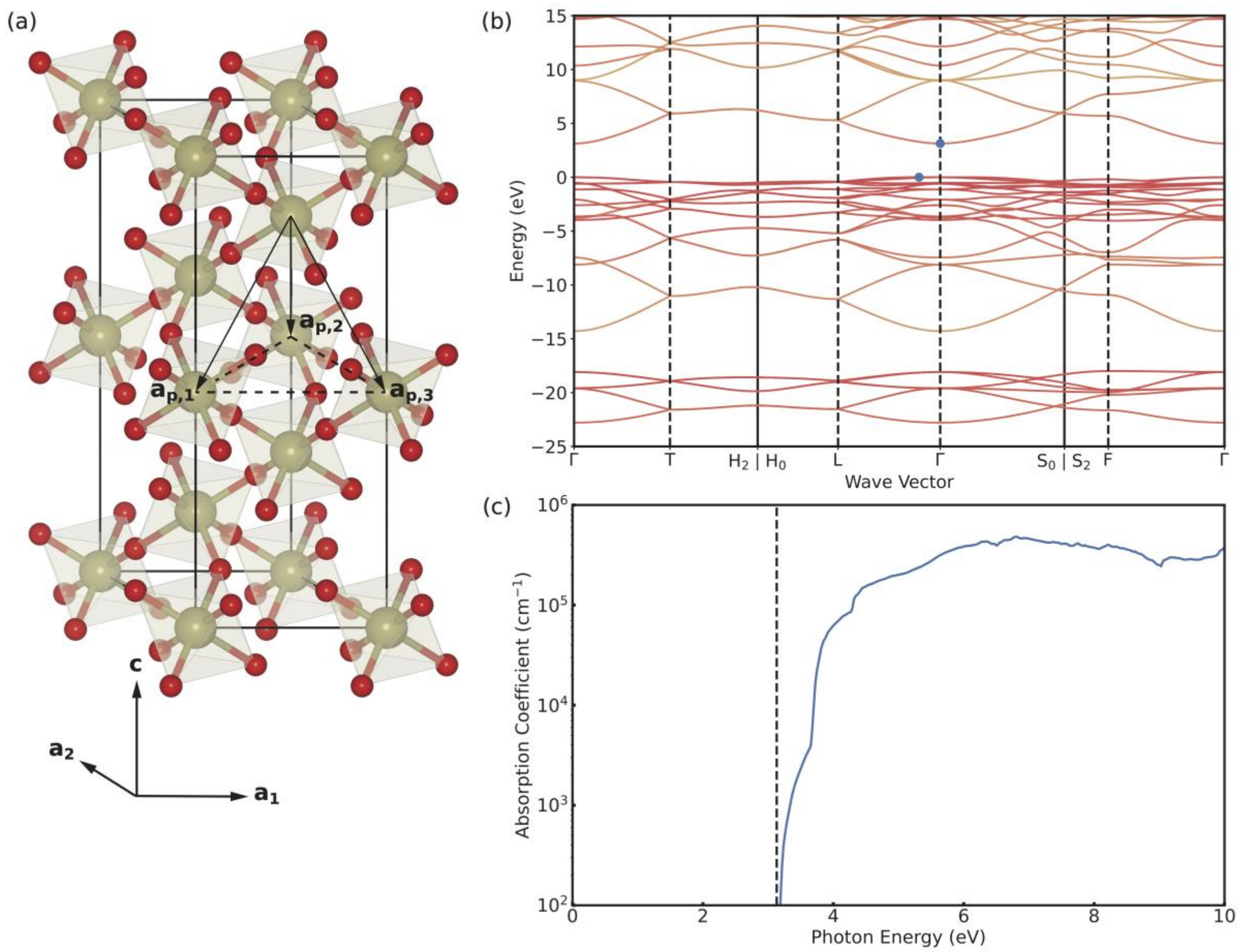

**Fig. 1.** (a) Conventional/hexagonal cell of $\beta$-$TeO_3$, with the primitive/rhombohedral cell dimensions embedded. Atoms in the khaki colour are Te and those in red are O. (b) Calculated band structure, with the VBM and CBM marked by blue dots. Te orbital components are plotted in the khaki colour and O orbital components in red; mixed characteristics by Te and O are coloured according to their respective compositions. The VBM is set to the zero energy. (c) Calculated optical absorption spectrum.

**Table 1.** Diagonal components (*xx* and *zz*) of the calculated electron and hole effective mass tensors ($m_e^*$ and $m_h^*$, respectively) and those of the dielectric tensors with electronic and ionic contributions ($\varepsilon_{ele}$ and $\varepsilon_{ion}$, respectively) of $\beta$-$TeO_3$. Effective masses are estimated for a carrier concentration of $10^{16}$ $cm^{-3}$ at 300 K, in reference to the free-electron rest mass $m_0$. Dielectric constants are in reference to vacuum permittivity.

| | Diagonal components of effective mass or dielectric tensors | |
|---|---|---|
| | *xx* | *zz* |
| $m_e^*/m_0$ | 0.64 | 0.44 |
| $m_h^*/m_0$ | 4.37 | 2.45 |
| $\varepsilon_{ele}$ | 3.83 | 4.09 |
| $\varepsilon_{ion}$ | 6.44 | 8.40 |

*3.2 Native defect formation and donor doping*

To discuss the carrier generation and compensation behaviour in $\beta$-$TeO_3$, native point defects, including vacancies, interstitials, and antisites, have been systematically investigated. Their formation energies are given as a function of the atomic chemical potentials and the Fermi level by definition as in Eq. (2). Here, two representative thermodynamic limits are considered, between which $\beta$-$TeO_3$ remains stable: O-rich conditions ($\Delta\mu_{\mathrm{Te}} = -3.43$ eV and $\Delta\mu_{\mathrm{O}} = 0$ eV) and O-poor conditions ($\Delta\mu_{\mathrm{Te}} = -2.85$ eV and $\Delta\mu_{\mathrm{O}} = -0.20$ eV), where $\Delta\mu_{\mathrm{Te}}$ and $\Delta\mu_{\mathrm{O}}$ are taken in reference to the total energies of elemental Te and an isolated $O_2$ molecule, respectively.

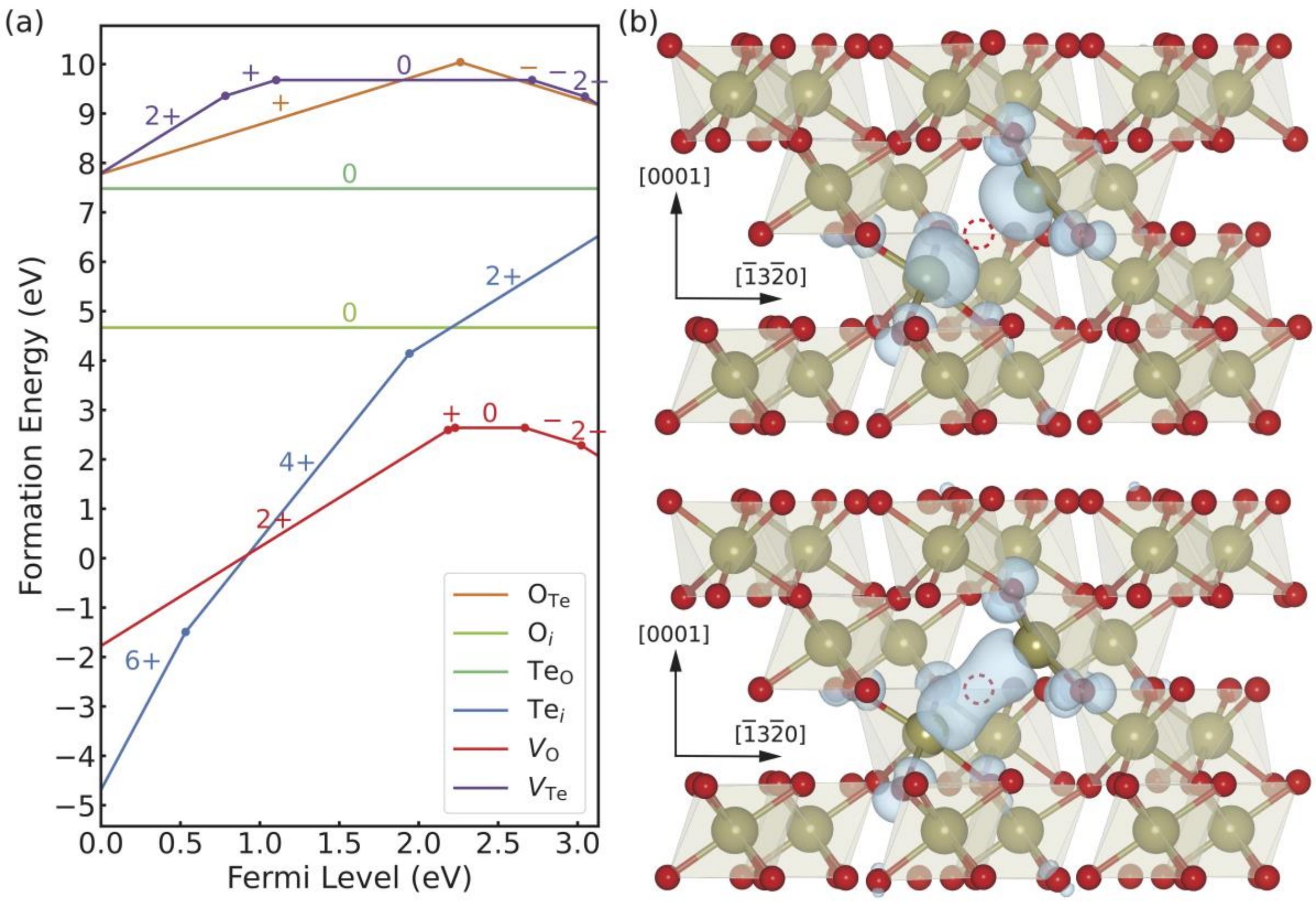

**Fig. 2.** (a) Formation energies of native point defects as a function of the Fermi level under the O-rich conditions. $V_X$ and $X_i$ ($X$ = Te or O) denote vacancies and interstitials, respectively. $X_Y$ are antisites with $X$ taking a $Y$ site ($X$, $Y$ = Te or O). The lower and upper limits of the Fermi level are the calculated VBM and CBM, respectively. For each defect type, only its formation energy in the most thermodynamically stable charge state is shown at a given Fermi level; the slope reflects the charge state in Eq. (2). The Fermi level positions, where the discontinuities in slope occur, correspond to charge transition levels, namely donor or acceptor levels. (b) Relaxed atomic configuration of $V_{\mathrm{O}}^{2-}$ together with the squared wavefunctions of two representative in-gap single-particle states, each containing one electron. They are enclosed by isosurfaces of 0.003 1/Å$^3$ coloured in cyan. The Te atoms are in the khaki colour and O in red. The O vacancy is represented by red dashed circles.

The formation energies of native point defects and the transition levels of their respective charge states, namely donor or acceptor levels, under O-rich conditions are shown in Fig. 2. The results for O-poor conditions, which differ only slightly due to the narrow thermodynamic stability window of $TeO_3$, are provided in Fig. S2 of the Supplementary materials. It can be observed in Fig. 2 that negative charge states are thermodynamically unfavoured near the VBM for all native defects, while their positive charge states all feature rather high formation energies near the CBM. Therefore, the formation of native shallow acceptors or donors, hence carrier generation in high density, is unlikely. In addition, the O vacancies exhibit a preference for negative charge states, namely $V_O^-$ and $V_O^{2-}$, when the Fermi level is close to the CBM. This arises from electron trapping around the vacancy and the adjacent Te sites that are displaced, as illustrated in Fig. 2b for the $V_O^{2-}$ configuration, with their corresponding single-particle states deep in the band gap. Similar electron trapping by O vacancies has been reported in other oxides such as amorphous $HfO_2$ [67]. The formation of such negatively charged O vacancies is a potential source of carrier compensation. However, their formation energies are as high as ~2 eV under O-rich conditions (also ~2 eV under O-poor conditions as in Fig. S2 of the Supplementary materials), indicating their low concentrations in $TeO_3$ under thermodynamic equilibrium at moderate temperatures and, consequently, their minor roles in carrier compensation.

The formation of STEs, including both one-electron and two-electron trapping cases, has been examined in accordance with the above Te-$V_O$-Te configuration on the corresponding Te-O-Te trimer in supercells without defects. No stabilised STE is observed across a series of candidate STE configurations (see Fig. S3 in the Supplementary materials for the atomic configuration of the supercell containing one additional electron to test the formation of an STE and the corresponding squared wavefunction of a single-particle state near the CBM, in which no obvious electron localisation is found). Therefore, electron transport via band conduction is likely in $\beta$-$TeO_3$.

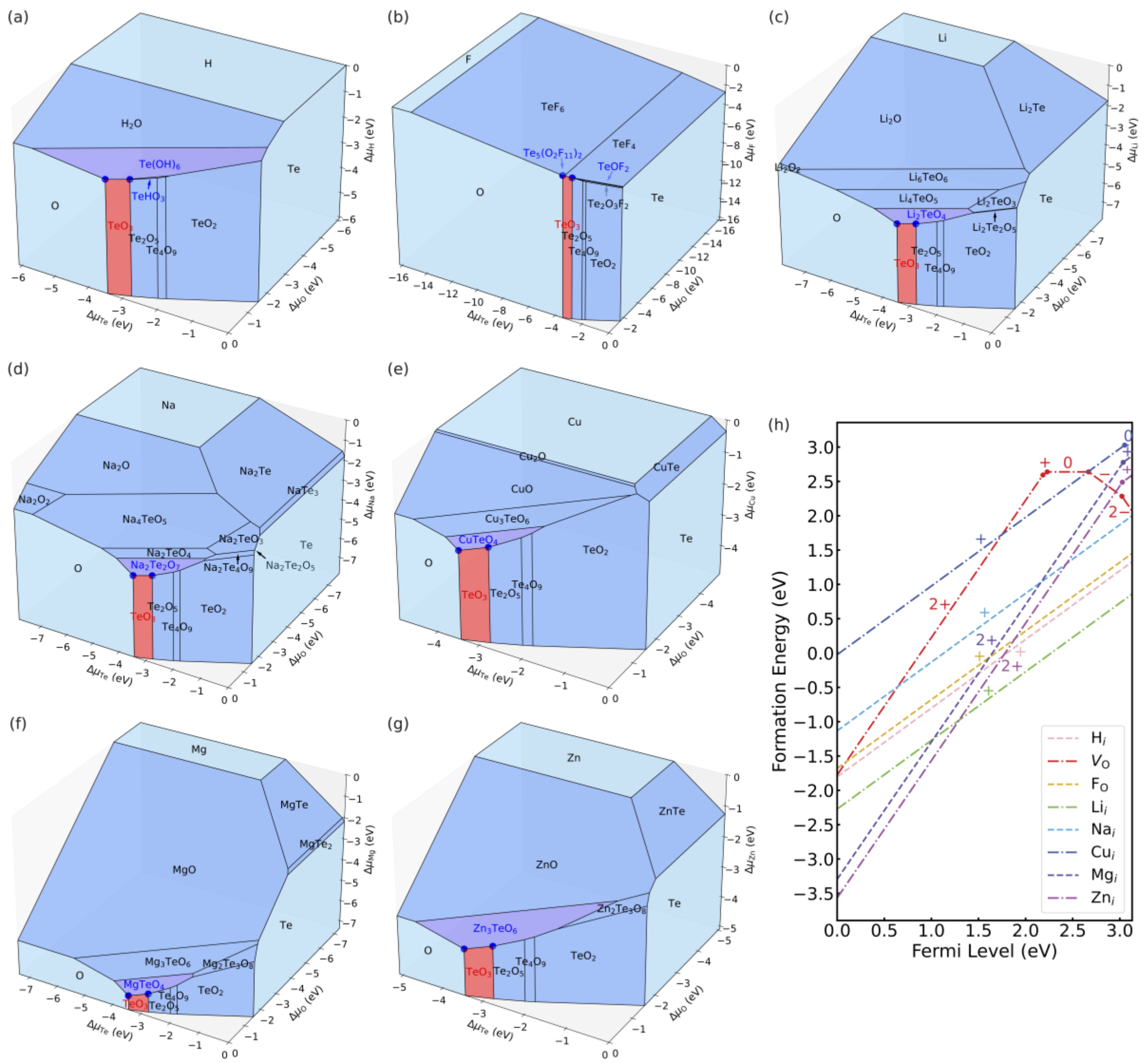

**Fig. 3.** Chemical potential diagrams for (a) Te-O-H, (b) Te-O-F, (c) Te-O-Li, (d) Te-O-Na, (e) Te-O-Cu, (f) Te-O-Mg, and (g) Te-O-Zn ternary systems. O-rich and -poor conditions adopted in this study are marked by blue dots, where the stable region of $TeO_3$ intersects with specific competing phases containing the impurity species, thereby corresponding to the solubility limits. (h) Formation energies of the impurities in $TeO_3$, together with that of the O vacancy, as a function of the Fermi level under O-rich conditions. $X_i$ and $X_O$ ($X$ = H, Li, Na, Cu, Mg, Zn, or F) denote the impurities at interstitial and O sites, respectively. Their formation energies and transition levels are shown similarly to those in Fig. 2a. The impurities located on other possible sites, in relevant charge states, with rather high formation energies are provided in Fig. S4 of the Supplementary materials.

Potential donor-type impurities in $TeO_3$ are further studied, considering both interstitial and substitutional configurations. The candidate donor species include H, Li, Na, Cu, Mg, Zn, and F. In particular, H impurities are often unintentionally incorporated into oxides and act as shallow donors [68]. As found in the chemical potential diagrams for Te-O-$X$ ($X$ = H, Li, Na,

Cu, Mg, Zn, and F) ternary systems in Fig. 3a–g, O-rich and O-poor conditions considered for native defects in undoped $TeO_3$ remain unchanged in terms of the Te and O chemical potentials. The chemical potential conditions of $X$ that correspond to their solubility limits as impurities in $TeO_3$ are considered here, whose values are summarised in Table S3 in the Supplementary materials.

The impurities taking interstitial, Te-, and O-substitutional sites have been investigated. As illustrated in Fig. 3h, some interstitial impurities ($H_i^+$, $Li_i^+$, and $Na_i^+$) and F substituting for O ($F_O^+$) emerge as shallow donors, exhibiting much lower formation energies than the native $V_O^{2-}$ at the Fermi level positions near the CBM and no in-gap transition levels. Interstitial Mg and Zn ($Mg_i$ and $Zn_i$) both feature 2+/+ transition levels at the Fermi level positions slightly below the CBM and comparable formation energies to the native $V_O^-$ and $V_O^{2-}$. All these impurities are expected to form hydrogenic donor states associated with the +/0 transition, although we cannot model such states in the present supercell approach for the reasons discussed above. Interstitial Cu ($Cu_i$) shows a +/0 transition level near the CBM but a much higher formation energy than the native $V_O^-$ and $V_O^{2-}$, and is less energetically favourable. Similar tendencies are observed for all the considered impurities under O-poor conditions (Fig. S5 in the Supplementary materials). These findings collectively suggest potential n-type conductivity in $TeO_3$ via shallow donors. In particular, H impurities unintentionally incorporated may account for the n-type conductivity of nominally undoped $TeO_3$, given that no native defects act as shallow donors according to our calculations. Te-substitutional impurities were identified as deep acceptors with high formation energies of above 3 eV across the Fermi level range within the band gap (see Fig. S4 of the Supplementary materials).

The overall behaviour of the native defects and dopants in $TeO_3$ towards the strong n-type preference is consistent with the band-alignment-based doping rule, where the low CBM and VBM with respect to the vacuum level, namely the high EA and IP, tend to facilitate n-type doping and hinder p-type doping [69–71]; the calculated band positions of $TeO_3$ are discussed below in Section 3.3.

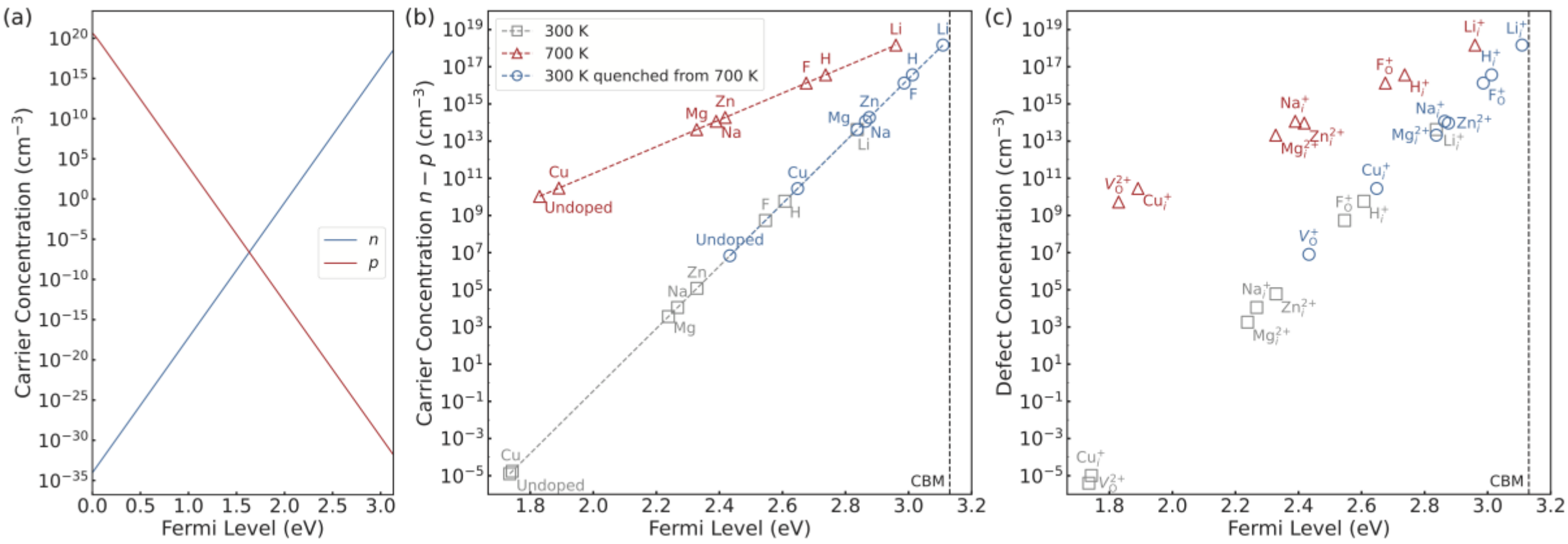

**Fig. 4.** (a) Electron ($n$) and hole ($p$) concentrations at 300 K against the Fermi level. (b) Equilibrium carrier concentrations ($n - p$) and Fermi levels at various temperatures in $\beta$-$TeO_3$, undoped or doped with different donor-type impurities (H, Li, Na, Cu, Mg, Zn, and F), obtained self-consistently with respect to the condition of charge neutrality. (c) Concentration of the most abundant defect with a non-zero charge state in each case corresponding to (b). (b) and (c) are under O-rich conditions.

The findings above are further verified by the equilibrium carrier concentrations and the corresponding Fermi levels in $\beta$-$TeO_3$. The relationship between the carrier concentrations and the Fermi level at 300 K is given in Fig. 4a and the net carrier concentrations in equilibrium due to donor doping at 300 K and 700 K in Fig. 4b. In the latter, the quenched defect concentrations from 700 K have been additionally considered for 300 K, where the total concentration of each defect type covering all its charge states is fixed from 700 K but the ratio of each charge state is redetermined according to the Boltzmann distribution at 300 K so that Eq. (5) is solved for the updated carrier concentrations. It is seen in Fig. 4b that the reported shallow donors with relatively low formation energies (namely, H, Li, and F) in their preferred defect configurations and charge states as in Fig. 4c, if quenched from 700 K, contribute to carrier concentrations of at least $10^{16}$ $cm^{-3}$ and Fermi levels above 3 eV that are sufficiently close to the calculated band gap 3.13 eV. Detailed carrier concentrations, defect concentrations, and Fermi levels under O-rich conditions are summarised in Tables S4–S6 of the Supplementary materials. We also note that the carrier concentration can be increased above the equilibrium values by supersaturating dopants under non-equilibrium crystal-growth or doping conditions.

### *3.3 Surface properties and band alignment*

The work functions of $TeO_3$ surfaces are predicted under the assumption that the Fermi level is located near the CBM, given the above results from the Fermi-level and carrier/impurity-concentration analyses for donor-doped $TeO_3$. In this situation, the EA provides a reasonable approximation to the work function. The work functions of both stoichiometric and O-deficient nonpolar surfaces are estimated accordingly. Surface energies of all investigated surfaces are provided in Fig. 5a. The band positions, evaluated by −EAs and −IPs respectively representing CBMs and VBMs with respect to the vacuum level, for all stoichiometric surfaces are summarised in Fig. 5b with the −EAs values for O-deficient surfaces. It is seen that the $(12\bar{3}2)$ surface features the lowest surface energy among all stoichiometric surfaces investigated, followed by $(11\bar{2}0)$ and $(10\bar{1}0)$, yielding EAs ranging from 6.86 to 8.46 eV. These values exceed the calculated EA of 6.3 eV from the stoichiometric (100) surface of $\alpha$-$MoO_3$ under an equivalent computational protocol except for the Tkatchenko–Scheffler dispersion correction [72] adopted for the relaxation of the layered structure. The calculated EA of $\alpha$-$MoO_3$ agrees reasonably with the reported experimental values in the range of 6.6–6.7 eV [8–10].

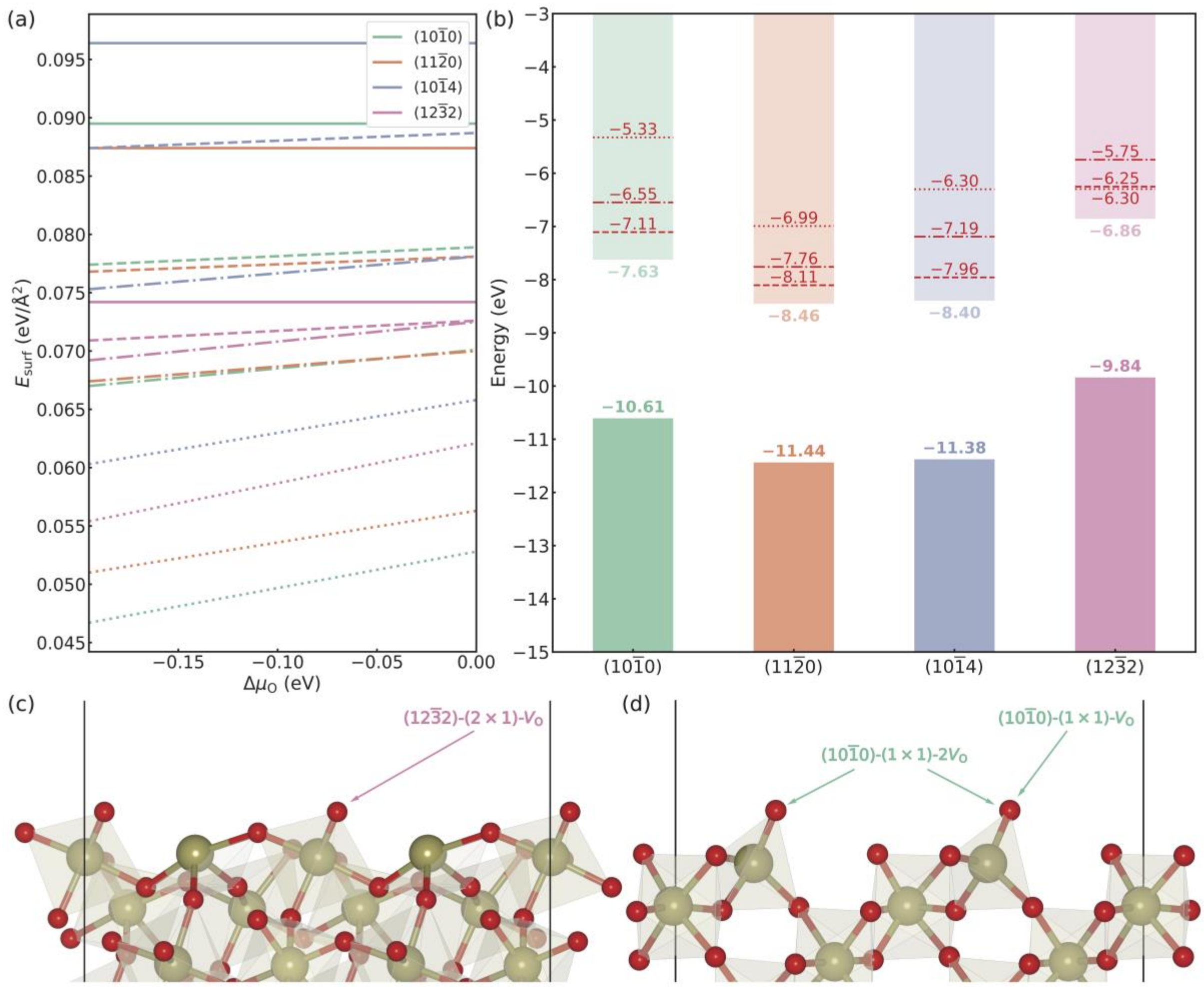

**Fig. 5.** (a) Surface energies of various surface orientations in $TeO_3$ with or without O vacancies. (b) Band alignment of all investigated surfaces in (a). The results from the most stable O-vacancy configurations with respect to distinct surface orientations are shown, corresponding to the cases with zero relative surface

energies in Fig. 6. In panel (a), solid, dashed, dash-dotted, and dotted lines represent stoichiometric surfaces, and surfaces with low, medium, and high O-vacancy concentrations, respectively. The latter three are respectively defined as in Fig. 6a–c. The upper edges of lower bars in (b) are aligned VBMs by −IPs and the lower edges of the upper ones are aligned CBMs by −EAs for stoichiometric surfaces; the respective −IP and −EA values are shown in the corresponding colours. The aligned CBMs of surfaces with low, medium, and high O-vacancy concentrations are given by dashed, dash-dotted, and dotted lines in red, respectively, with corresponding −EA values provided. (c) Side view of the $(12\bar{3}2)$ surface in 2×1 units and (d) that of the $(10\bar{1}0)$ surface in the 1×1 unit. The most energetically preferred positions of O vacancies are indicated by corresponding arrows.

The O vacancies are first introduced to surfaces of increased areas to approximately model the O-deficient surfaces with low vacancy concentrations. Such increased surface areas are denoted by 2×1 or 2×2 units of the in-plane size of its corresponding stoichiometric surface. Different vacancy sites defined by their relative distances from the surface (vacuum) are examined as shown in Fig. 6a, so that the surface energies and band positions of the most energetically favourable vacancy configurations are reported in Figs. 5a and 5b. Surfaces of low O-vacancy concentrations are found to be more energetically preferred with respect to their corresponding stoichiometric surfaces for all surface orientations of interest, in particular with O vacancies preferentially forming near the vacuum region. The surface O vacancies reduce the calculated EAs (hence the estimated work functions) for all investigated surfaces by about 0.5 eV. The $(12\bar{3}2)$ surface is still the most stable among all investigated surfaces with low O-vacancy concentrations, whose estimated work function is reduced to 6.25 eV.

The origin of high theoretical work functions in $TeO_3$ appears to be similar to $\alpha$-$MoO_3$ [13]; the Te-O bonds towards the vacuum constitute significant dipole layers (pointing to the vacuum) for the stoichiometric surfaces. Such surface Te-O bonds yield 1-fold coordinated (undercoordinated) O sites as shown in Figs. 5c and 5d, and there are multiple of them per unit surface. Notably in Fig. 6a, O vacancies taking 1-fold coordinated surface sites, especially those close to the vacuum, are in general more energetically favoured. Surface O vacancies break the dipole-layer arrangements due to missing Te-O bonds, reducing the surface dipole contribution, hence the work functions and EA values.

Surfaces of larger O deficiency are further investigated: the low O-vacancy concentrations in Fig. 6a are doubled in Fig. 6b (denoted by medium concentrations) and further doubled in Fig. 6c (high concentrations). Medium surface O-vacancy concentrations are achieved by reducing the in-plane surface areas by 50% from those with low concentrations, to which different vacancy sites are equivalently defined. Fig. 5d shows the atomic structure of the (10$\bar{1}$0)-(1×1) surface. The (10$\bar{1}$0) surface of a high O-vacancy concentration is constructed by removing the other O atom taking an equivalent surface site, as labelled by (10$\bar{1}$0)-(1×1)-2$V_O$ in Fig. 5d pointing to the O atoms removed to generate vacancies, with no other vacancy configurations considered. Those of (11$\bar{2}$0) and (10$\bar{1}$4) orientations are simply made by further reducing the in-plane area, considering the vacancy sites at different distances from the surfaces. The (12$\bar{3}$2) surfaces of a high vacancy concentration contain two O vacancies per 1×1 unit, with the first taking the closest O site to the vacuum but the other occupying the second, third, or fourth nearest site to the vacuum, corresponding to the three cases/distances in Fig. 6c labelled by (12$\bar{3}$2)-(1×1)-2$V_O$.

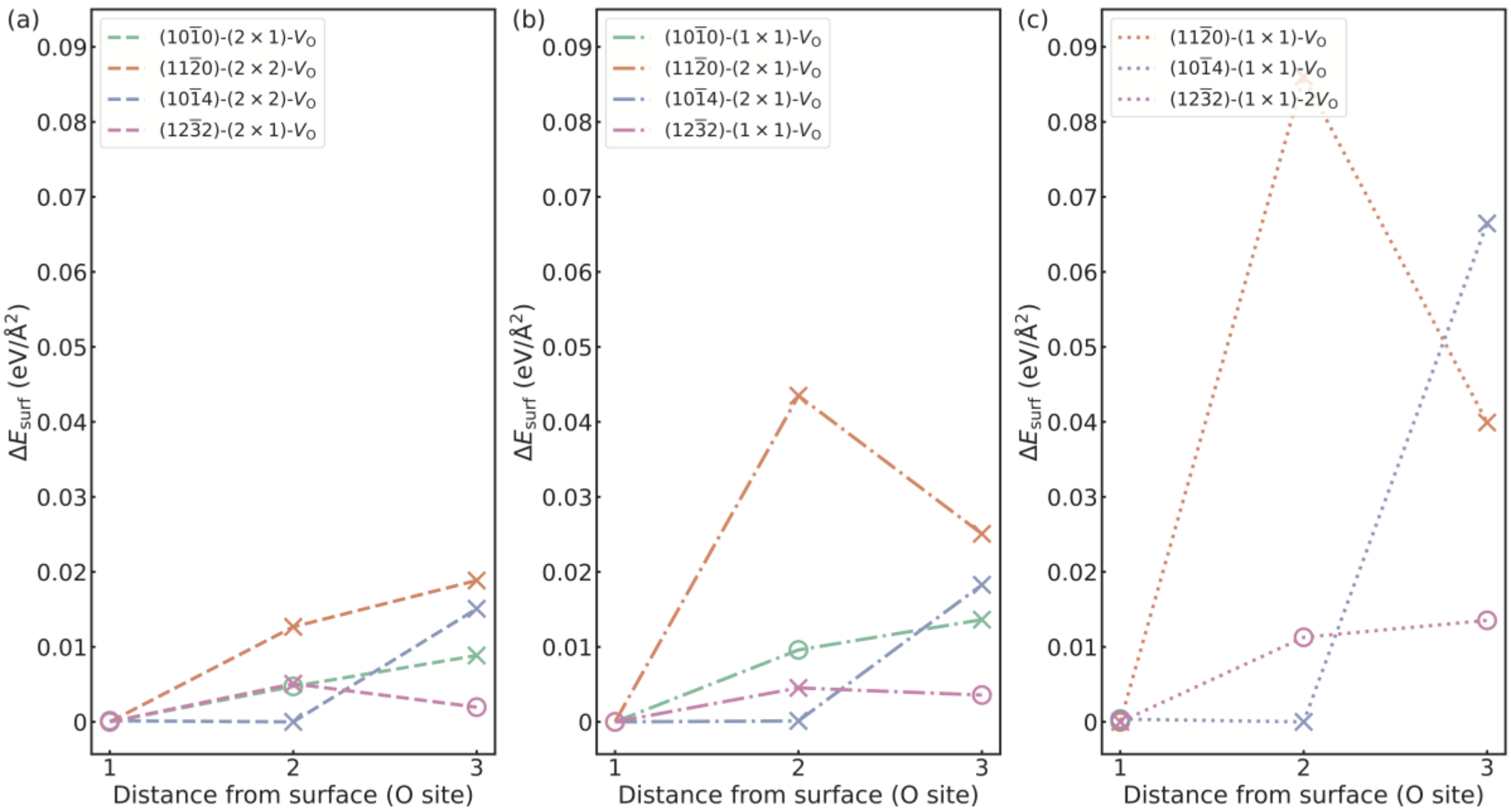

**Fig. 6.** Relative surface energy against the distance at which the O vacancy is located from the surface: (a) low, (b) medium, and (c) high vacancy concentrations. The distance is qualitatively represented by the first, second, and third nearest O sites relative to the vacuum region. Each model contains only one O vacancy per unit surface, labelled by -$V_O$, except for (12$\bar{3}$2)-(1×1)-2$V_O$ in (c), constructed from (12$\bar{3}$2)-(1×1)-$V_O$ in (b), with one extra O vacancy created per unit surface, whose distance from the surface is plotted. In each subplot, relative energies are reported for respective surface orientations and referenced to the lowest-energy vacancy configuration of each orientation. Vacancies occupying 1-fold coordinated O sites are indicated by circles and 2-fold coordinated O sites by crosses.

The surface energies and band positions of medium and high O-vacancy concentrations are plotted in Figs. 5a and 5b, along with the stoichiometric and low O-vacancy concentration cases. The (10$\bar{1}$0) surfaces become the most energetically preferred when more surface O vacancies are present. The calculated EAs are further reduced, e.g. 5.33 eV for $(10\bar{1}0)$, due to missing surface Te-O bonds that account for the surface dipole contribution to the vacuum. These results indicate the potential detrimental effect of O vacancies in $TeO_3$ on not only its n-type conductivity but also its work functions. Therefore, synthesising and/or treating $TeO_3$ under the conditions of a high O chemical potential is strongly recommended, increasing the O chemical potential even to a level beyond the formation limit of molecular $O_2$, such as by using oxygen plasma treatments or ozone precursors. The O chemical potential necessary to maintain surface stoichiometry is further calculated by considering surface energies for $\Delta\mu_O >$ 0 eV (see Fig. S6 of Supplementary materials) and the major transition from the O-deficient $(10\bar{1}0)$ surface to the stoichiometric $(12\bar{3}2)$ surface is observed at $\Delta\mu_O$ = 0.68 eV. Note that the stoichiometric $(12\bar{3}2)$ surface yields an estimated work function of ~6.9 eV as in Fig. 5b, while the highly O-deficient $(11\bar{2}0)$ surface, with a very similar surface energy to the most stable $(10\bar{1}0)$ case, still maintains a rather high work function of ~7 eV. Therefore, the target ultrahigh work functions should be achieved by considering specific surface orientations in parallel with elevating the O chemical potential.

**4. Conclusions**

To summarise, $\beta$-$TeO_3$ demonstrates significant potential as an n-type wide-gap semiconductor through impurity doping with interstitial H, Li, Na, Cu, Mg, and Zn or substitutional F on O sites, all as shallow donors. The formation of O vacancies can give rise to electron trapping, acting as a source of electron carrier compensation, but their high formation energies indicate that this behaviour can be sufficiently suppressed. Ultrahigh work functions have been predicted across a variety of nonpolar surfaces of $\beta$-$TeO_3$, although these work functions can be impacted by O vacancies formed near the surfaces. The exceptional electronic properties of $\beta$-$TeO_3$ characterised by optical transparency, n-type conductivity, and in particular ultrahigh work functions, revealed by various point-defect and surface calculations, can therefore be achieved through cautious control over the formation of surface O vacancies. The present findings offer valuable insights into defect and surface engineering in $\beta$-$TeO_3$ for its practical applications in electronic and opto-electronic devices requiring transparent semiconductors with high work functions.

**Data availability**

The data that support the findings of this study are available from the corresponding authors upon reasonable request.

**CRediT authorship contribution statement**

**Tianwei Wang:** Conceptualisation, Data curation, Formal analysis, Investigation, Methodology, Project administration, Software, Validation, Visualisation, Writing - original draft, Writing - review & editing. **Teruya Nagafuji:** Data curation, Formal analysis, Investigation, Methodology, Software, Writing - review & editing. **Kota Hanzawa** and **Hidenori Hiramatsu:** Project administration, Validation, Writing - review & editing. **Fumiyasu Oba:** Conceptualisation, Funding acquisition, Methodology, Project administration, Resources, Supervision, Validation, Writing - review & editing.

**Declaration of competing interest**

The authors have a pending patent application with KISTEC and Institute of Science Tokyo, which is related to the findings in this article, and declare no other competing financial interests or personal relationships that could have appeared to influence the work reported in this paper.

**Acknowledgements**

This work was supported by the KISTEC Project, JSPS KAKENHI (grant numbers JP24H00376, JP24K21683, and JP26K22583), and MEXT Data Creation and Utilization Type Material Research and Development Project (grant number JPMXP1122683430). The calculations were performed using the Large-scale Supercomputer Systems of ACCMS at Kyoto University, the computational resources offered under the category of General Projects by the Research Institute for Information Technology at Kyushu University, and the TSUBAME4.0 supercomputer at Institute of Science Tokyo. The authors also acknowledge the Center for Computational Materials Science, Institute for Materials Research, Tohoku University for the use of MASAMUNE-II (project number 2512SC0017).

## References

[1] H.B. Michaelson, The work function of the elements and its periodicity, J. Appl. Phys. 48 (1977) 4729–4733. https://doi.org/10.1063/1.323539.

[2] Y.S. Park, E. Kim, B. Hong, J. Lee, Characteristics of ITO films with oxygen plasma treatment for thin film solar cell applications, Mater. Res. Bull. 48 (2013) 5115–5120. https://doi.org/10.1016/j.materresbull.2013.07.026.

[3] X. Rong, J. Han, C. Xu, B. Ma, L. Jiang, D. Ma, R. Xing, L. Shen, L. Duan, Y. Deng, Y. Geng, G. Dong, Nitrogenous Interlayers for ITO S/D Electrodes in N-Type Organic Thin Film Transistors, Adv. Mater. Interfaces 10 (2023) 2202059. https://doi.org/10.1002/admi.202202059.

[4] H. Peelaers, M.L. Chabinyc, C.G. Van de Walle, Controlling n-Type Doping in $MoO_3$, Chem. Mater. 29 (2017) 2563–2567. https://doi.org/10.1021/acs.chemmater.6b04479.

[5] J. Meyer, R. Khalandovsky, P. Görrn, A. Kahn, $MoO_3$ Films Spin-Coated from a Nanoparticle Suspension for Efficient Hole-Injection in Organic Electronics, Adv. Mater. 23 (2011) 70–73. https://doi.org/10.1002/adma.201003065.

[6] J. Meyer, S. Hamwi, M. Kröger, W. Kowalsky, T. Riedl, A. Kahn, Transition Metal Oxides for Organic Electronics: Energetics, Device Physics and Applications, Adv. Mater. 24 (2012) 5408–5427. https://doi.org/10.1002/adma.201201630.

[7] M.T. Greiner, M.G. Helander, W.M. Tang, Z. Bin Wang, J. Qiu, Z.H. Lu, Universal energy-level alignment of molecules on metal oxides, Nat. Mater. 11 (2012) 76–81. https://doi.org/10.1038/nmat3159.

[8] P. Schulz, J.O. Tiepelt, J.A. Christians, I. Levine, E. Edri, E.M. Sanehira, G. Hodes, D. Cahen, A. Kahn, High-Work-Function Molybdenum Oxide Hole Extraction Contacts in Hybrid Organic–Inorganic Perovskite Solar Cells, ACS Appl. Mater. Interfaces 8 (2016) 31491–31499. https://doi.org/10.1021/acsami.6b10898.

[9] M. Kröger, S. Hamwi, J. Meyer, T. Riedl, W. Kowalsky, A. Kahn, Role of the deep-lying electronic states of $MoO_3$ in the enhancement of hole-injection in organic thin films, Appl. Phys. Lett. 95 (2009) 123301. https://doi.org/10.1063/1.3231928.

[10] A.L.F. Cauduro, R. Dos Reis, G. Chen, A.K. Schmid, C. Méthivier, H.G. Rubahn, L. Bossard-Giannesini, H. Cruguel, N. Witkowski, M. Madsen, Crystalline Molybdenum Oxide Thin-Films for Application as Interfacial Layers in Optoelectronic Devices, ACS Appl. Mater. Interfaces 9 (2017) 7717–7724. https://doi.org/10.1021/acsami.6b14228.

[11] A.L.F. Cauduro, R. dos Reis, G. Chen, A.K. Schmid, H.-G. Rubahn, M. Madsen, Work function mapping of $MoO_x$ thin-films for application in electronic devices, Ultramicroscopy 183 (2017) 99–103. https://doi.org/10.1016/j.ultramic.2017.03.025.

[12] C. Lattyak, K. Gehrke, M. Vehse, Layer-Thickness-Dependent Work Function of $MoS_2$ on Metal and Metal Oxide Substrates, J. Phys. Chem. C 126 (2022) 13929–13935. https://doi.org/10.1021/acs.jpcc.2c03268.

[13] Y. Guo, J. Robertson, Origin of the high work function and high conductivity of $MoO_3$, Appl. Phys. Lett. 105 (2014) 222110. https://doi.org/10.1063/1.4903538.

[14] Y. Hinuma, Y. Kumagai, I. Tanaka, F. Oba, Effects of composition, crystal structure, and surface orientation on band alignment of divalent metal oxides: A first-principles study, Phys. Rev. Mater. 2 (2018) 124603. https://doi.org/10.1103/PhysRevMaterials.2.124603.

[15] Y. Hinuma, T. Gake, F. Oba, Band alignment at surfaces and heterointerfaces of $Al_2O_3$, $Ga_2O_3$, $In_2O_3$, and related group-III oxide polymorphs: A first-principles study, Phys. Rev. Mater. 3 (2019) 084605. https://doi.org/10.1103/PhysRevMaterials.3.084605.

[16] S. Kiyohara, Y. Hinuma, F. Oba, Band Alignment of Oxides by Learnable Structural-Descriptor-Aided Neural Network and Transfer Learning, J. Am. Chem. Soc. 146 (2024) 9697–9708. https://doi.org/10.1021/jacs.3c13574.

[17] M. Dušek, J. Loub, X-ray Powder Diffraction Data and Structure Refinement of $TeO_3$, Powder Diffr. 3 (1988) 175–176. https://doi.org/10.1017/S0885715600032942.

[18] J. Loub, Kristallines Tellurtrioxid $TeO_3$, Z. Anorg. Allg. Chem. 362 (1968) 98–107. https://doi.org/10.1002/zaac.19683620116.

[19] J. Loub, J. Rosický, Über $\alpha$-$TeO_3$, $Te_2O_5$ und die paramagnetischen Produkte der thermischen Zersetzung von $H_6TeO_6$, Z. Anorg. Allg. Chem. 365 (1969) 308–314. https://doi.org/10.1002/zaac.19693650511.

[20] J.C.J. Bart, A. Bossi, P. Perissinoto, A. Castellan, N. Giordano, Some observations on the thermochemistry of telluric acid, J. Therm. Anal. 8 (1975) 313–327. https://doi.org/10.1007/BF01904009.

[21] J. Loub, Vibrational spectra of tellurium trioxide and ditelluryl sulfate, Collect. Czechoslov. Chem. Commun. 42 (1977) 960–966. https://doi.org/10.1135/cccc19770960.

[22] M.A.K. Ahmed, H. Fjellvåg, A. Kjekshus, Synthesis, structure and thermal stability of tellurium oxides and oxide sulfate formed from reactions in refluxing sulfuric acid, J. Chem. Soc., Dalton Trans. (2000) 4542–4549. https://doi.org/10.1039/b005688j.

[23] V.P. Itkin, C.B. Alcock, The O-Te (oxygen-tellurium) system, J. Phase Equilib. 17 (1996) 533–538. https://doi.org/10.1007/BF02666000.

[24] S. Gossé, Tellurium Oxides: Thermodynamics and Phase Relations in the Te–O System, J. Phase Equilib. Diffus. 46 (2025) 20–30. https://doi.org/10.1007/s11669-025-01175-6.

[25] N. Dewan, K. Sreenivas, V. Gupta, Influence of $\gamma$-radiation doses on the properties of $TeO_x$: ($x$=2–3) thin film, J. Appl. Phys. 102 (2007) 044906. https://doi.org/10.1063/1.2769778.

[26] S. Bang, C. Lee, D. Choi, D.Y. Park, D.H. Kim, D. Lee, D. Yi, J. Song, S.J. Yun, D. Kim, M.S. Jeong, High Performance P-Channel Transistor Based on Amorphous Tellurium Trioxide, Adv. Mater. 37 (2025) 2504948. https://doi.org/10.1002/adma.202504948.

[27] P.E. Blöchl, Projector augmented-wave method, Phys. Rev. B 50 (1994) 17953–17979. https://doi.org/10.1103/PhysRevB.50.17953.

[28] A. Seidl, A. Görling, P. Vogl, J.A. Majewski, M. Levy, Generalized Kohn-Sham schemes and the band-gap problem, Phys. Rev. B 53 (1996) 3764–3774. https://doi.org/10.1103/PhysRevB.53.3764.

[29] G. Kresse, J. Hafner, Ab initio molecular dynamics for liquid metals, Phys. Rev. B 47 (1993) 558–561. https://doi.org/10.1103/PhysRevB.47.558.

[30] G. Kresse, J. Furthmüller, Efficient iterative schemes for ab initio total-energy calculations using a plane-wave basis set, Phys. Rev. B 54 (1996) 11169–11186. https://doi.org/10.1103/PhysRevB.54.11169.

[31] G. Kresse, D. Joubert, From ultrasoft pseudopotentials to the projector augmented-wave method, Phys. Rev. B 59 (1999) 1758–1775. https://doi.org/10.1103/PhysRevB.59.1758.

[32] J. Heyd, G.E. Scuseria, M. Ernzerhof, Hybrid functionals based on a screened Coulomb potential, J. Chem. Phys. 118 (2003) 8207–8215. https://doi.org/10.1063/1.1564060.

[33] J. Heyd, G.E. Scuseria, Assessment and validation of a screened Coulomb hybrid density functional, J. Chem. Phys. 120 (2004) 7274–7280. https://doi.org/10.1063/1.1668634.

[34] A. V. Krukau, O.A. Vydrov, A.F. Izmaylov, G.E. Scuseria, Influence of the exchange screening parameter on the performance of screened hybrid functionals, J. Chem. Phys. 125 (2006) 224106. https://doi.org/10.1063/1.2404663.

[35] K. Momma, F. Izumi, VESTA 3 for three-dimensional visualization of crystal, volumetric and morphology data, J. Appl. Crystallogr. 44 (2011) 1272–1276. https://doi.org/10.1107/S0021889811038970.

[36] R.W. Nunes, X. Gonze, Berry-phase treatment of the homogeneous electric field perturbation in insulators, Phys. Rev. B 63 (2001) 155107. https://doi.org/10.1103/PhysRevB.63.155107.

[37] I. Souza, J. Íñiguez, D. Vanderbilt, First-Principles Approach to Insulators in Finite Electric Fields, Phys. Rev. Lett. 89 (2002) 117602. https://doi.org/10.1103/PhysRevLett.89.117602.

[38] M. Gajdoš, K. Hummer, G. Kresse, J. Furthmüller, F. Bechstedt, Linear optical properties in the projector-augmented wave methodology, Phys. Rev. B 73 (2006) 045112. https://doi.org/10.1103/PhysRevB.73.045112.

[39] G.K.H. Madsen, J. Carrete, M.J. Verstraete, BoltzTraP2, a program for interpolating band structures and calculating semi-classical transport coefficients, Comput. Phys. Commun. 231 (2018) 140–145. https://doi.org/10.1016/j.cpc.2018.05.010.

[40] S.P. Ong, W.D. Richards, A. Jain, G. Hautier, M. Kocher, S. Cholia, D. Gunter, V.L. Chevrier, K.A. Persson, G. Ceder, Python Materials Genomics (pymatgen): A robust, open-source python library for materials analysis, Comput. Mater. Sci. 68 (2013) 314–319. https://doi.org/10.1016/j.commatsci.2012.10.028.

[41] Y. Kumagai, N. Tsunoda, A. Takahashi, F. Oba, Insights into oxygen vacancies from high-throughput first-principles calculations, Phys. Rev. Mater. 5 (2021) 123803. https://doi.org/10.1103/PhysRevMaterials.5.123803.

[42] P.Y. Yu, M. Cardona, Fundamentals of Semiconductors, Springer Berlin Heidelberg, Berlin, Heidelberg, 2010. https://doi.org/10.1007/978-3-642-00710-1.

[43] I. Mosquera-Lois, S.R. Kavanagh, A. Walsh, D.O. Scanlon, ShakeNBreak: Navigating the defect configurational landscape, J. Open Source Softw. 7 (2022) 4817. https://doi.org/10.21105/joss.04817.

[44] I. Mosquera-Lois, S.R. Kavanagh, A. Walsh, D.O. Scanlon, Identifying the ground state structures of point defects in solids, NPJ Comput. Mater. 9 (2023) 25. https://doi.org/10.1038/s41524-023-00973-1.

[45] N. Tsunoda, Y. Kumagai, F. Oba, Recommendation of interstitial hydrogen positions in metal oxides, Comput. Mater. Sci. 203 (2022) 111068. https://doi.org/10.1016/j.commatsci.2021.111068.

[46] C. Freysoldt, J. Neugebauer, C.G. Van de Walle, Fully Ab Initio Finite-Size Corrections for Charged-Defect Supercell Calculations, Phys. Rev. Lett. 102 (2009) 016402. https://doi.org/10.1103/PhysRevLett.102.016402.

[47] Y. Kumagai, F. Oba, Electrostatics-based finite-size corrections for first-principles point defect calculations, Phys. Rev. B 89 (2014) 195205. https://doi.org/10.1103/PhysRevB.89.195205.

[48] A. Jain, S.P. Ong, G. Hautier, W. Chen, W.D. Richards, S. Dacek, S. Cholia, D. Gunter, D. Skinner, G. Ceder, K.A. Persson, Commentary: The Materials Project: A materials genome approach to accelerating materials innovation, APL Mater. 1 (2013) 011002. https://doi.org/10.1063/1.4812323.

[49] A. Jain, G. Hautier, S.P. Ong, C.J. Moore, C.C. Fischer, K.A. Persson, G. Ceder, Formation enthalpies by mixing GGA and GGA+*U* calculations, Phys. Rev. B 84 (2011) 045115. https://doi.org/10.1103/PhysRevB.84.045115.

[50] M. Aykol, S.S. Dwaraknath, W. Sun, K.A. Persson, Thermodynamic limit for synthesis of metastable inorganic materials, Sci. Adv. 4 (2018) eaaq0148. https://doi.org/10.1126/sciadv.aaq0148.

[51] A. Wang, R. Kingsbury, M. McDermott, M. Horton, A. Jain, S.P. Ong, S. Dwaraknath, K.A. Persson, A framework for quantifying uncertainty in DFT energy corrections, Sci. Rep. 11 (2021) 15496. https://doi.org/10.1038/s41598-021-94550-5.

[52] C.J. Bartel, Review of computational approaches to predict the thermodynamic stability of inorganic solids, J. Mater. Sci. 57 (2022) 10475–10498. https://doi.org/10.1007/s10853-022-06915-4.

[53] J.P. Perdew, A. Ruzsinszky, G.I. Csonka, O.A. Vydrov, G.E. Scuseria, L.A. Constantin, X. Zhou, K. Burke, Restoring the Density-Gradient Expansion for Exchange in Solids and Surfaces, Phys. Rev. Lett. 100 (2008) 136406. https://doi.org/10.1103/PhysRevLett.100.136406.

[54] F. Oba, Y. Kumagai, Design and exploration of semiconductors from first principles: A review of recent advances, Appl. Phys. Express 11 (2018) 060101. https://doi.org/10.7567/APEX.11.060101.

[55] Y. Hinuma, A. Grüneis, G. Kresse, F. Oba, Band alignment of semiconductors from density-functional theory and many-body perturbation theory, Phys. Rev. B 90 (2014) 155405. https://doi.org/10.1103/PhysRevB.90.155405.

[56] T. Wang, N. Sato, F. Oba, Surface reconstruction and band alignment of non-metallic spinel oxides from first principles, Acta Mater. 293 (2025) 121034. https://doi.org/10.1016/j.actamat.2025.121034.

[57] T. Wang, F. Oba, Slab surface passivation and its application to band offset calculations for polar heterointerfaces of zinc-blende semiconductors, Phys. Rev. Mater. 7 (2023) 084602. https://doi.org/10.1103/PhysRevMaterials.7.084602.

[58] E.M. Roginskii, V.G. Kuznetsov, M.B. Smirnov, O. Noguera, J.R. Duclère, M. Colas, O. Masson, P. Thomas, Comparative Analysis of the Electronic Structure and Nonlinear Optical Susceptibility of $\alpha$-$TeO_2$ and $\beta$-$TeO_3$ Crystals, J. Phys. Chem. C 121 (2017) 12365–12374. https://doi.org/10.1021/acs.jpcc.7b01819.

[59] W.S. Baer, Faraday Rotation in ZnO: Determination of the Electron Effective Mass, Phys. Rev. 154 (1967) 785–789. https://doi.org/10.1103/PhysRev.154.785.

[60] N. Preissler, O. Bierwagen, A.T. Ramu, J.S. Speck, Electrical transport, electrothermal transport, and effective electron mass in single-crystalline $In_2O_3$ films, Phys. Rev. B 88 (2013) 085305. https://doi.org/10.1103/PhysRevB.88.085305.

[61] M. Nagasawa, S. Shionoya, S. Makishima, Electron Effective Mass of $SnO_2$, J. Phys. Soc. Japan 20 (1965) 1093. https://doi.org/10.1143/JPSJ.20.1093.

[62] F. Ricci, W. Chen, U. Aydemir, G.J. Snyder, G.-M. Rignanese, A. Jain, G. Hautier, An ab initio electronic transport database for inorganic materials, Sci. Data 4 (2017) 170085. https://doi.org/10.1038/sdata.2017.85.

[63] C. Jagannath, Z.W. Grabowski, A.K. Ramdas, Linewidths of the electronic excitation spectra of donors in silicon, Phys. Rev. B 23 (1981) 2082–2098. https://doi.org/10.1103/PhysRevB.23.2082.

[64] B.K. Meyer, J. Sann, D.M. Hofmann, C. Neumann, A. Zeuner, Shallow donors and acceptors in ZnO, Semicond. Sci. Technol. 20 (2005) S62–S66. https://doi.org/10.1088/0268-1242/20/4/008.

[65] J.S. Smith, A. Budi, M.C. Per, N. Vogt, D.W. Drumm, L.C.L. Hollenberg, J.H. Cole, S.P. Russo, Ab initio calculation of energy levels for phosphorus donors in silicon, Sci. Rep. 7 (2017) 6010. https://doi.org/10.1038/s41598-017-06296-8.

[66] G. Zhang, A. Canning, N. Grønbech-Jensen, S. Derenzo, L.-W. Wang, Shallow Impurity Level Calculations in Semiconductors Using Ab Initio Methods, Phys. Rev. Lett. 110 (2013) 166404. https://doi.org/10.1103/PhysRevLett.110.166404.

[67] J.L. Gavartin, D. Muñoz Ramo, A.L. Shluger, G. Bersuker, B.H. Lee, Negative oxygen vacancies in HfO2 as charge traps in high-k stacks, Appl. Phys. Lett. 89 (2006) 082908. https://doi.org/10.1063/1.2236466.

[68] A. Alkauskas, M.D. McCluskey, C.G. Van De Walle, Tutorial: Defects in semiconductors - Combining experiment and theory, J. Appl. Phys. 119 (2016). https://doi.org/10.1063/1.4948245.

[69] S.B. Zhang, S.-H. Wei, A. Zunger, Microscopic Origin of the Phenomenological Equilibrium "Doping Limit Rule" in n-Type III-V Semiconductors, Phys. Rev. Lett. 84 (2000) 1232–1235. https://doi.org/10.1103/PhysRevLett.84.1232.

[70] H. Hosono, Transparent Oxide Semiconductors: Fundamentals and Recent Progress, in: Transparent Electronics: From Synthesis to Applications, Wiley, 2010: pp. 31–59. https://doi.org/10.1002/9780470710609.ch2.

[71] J. Robertson, S.J. Clark, Limits to doping in oxides, Phys. Rev. B 83 (2011) 075205. https://doi.org/10.1103/PhysRevB.83.075205.

[72] A. Tkatchenko, M. Scheffler, Accurate Molecular Van Der Waals Interactions from Ground-State Electron Density and Free-Atom Reference Data, Phys. Rev. Lett. 102 (2009) 073005. https://doi.org/10.1103/PhysRevLett.102.073005.

**Supplementary materials for "Tellurium trioxide as an n-type wide-gap semiconductor with ultrahigh work functions from first principles"**

Tianwei Wang [a,b,*], Teruya Nagafuji [b], Kota Hanzawa [b], Hidenori Hiramatsu [b,c], and Fumiyasu Oba [a,b,c,*]

[a.] Kanagawa Institute of Industrial Science and Technology (KISTEC), Ebina 243-0435, Japan
[b.] Materials and Structures Laboratory, Institute of Integrated Research, Institute of Science Tokyo, Yokohama 226-8501, Japan
[c.] MDX Research Center for Element Strategy, Institute of Integrated Research, Institute of Science Tokyo, Yokohama 226-8501, Japan
[*] Corresponding authors.

E-mail addresses: wang.t@msl.iir.isct.ac.jp (T. Wang), oba.f@msl.iir.isct.ac.jp (F. Oba)

The supplementary materials contain the tables and figures that support the discussion presented in the associated paper.

## PAW datasets

**Table S1.** Valence electrons and core radii of the PAW datasets used in this study.

| | Valence-electron configuration | PAW core radius (Å) |
|---|---|---|
| Te | $5s^2\ 5p^4$ | 1.22 |
| O | $2s^2\ 2p^4$ | 0.80 |
| H | $1s^1$ | 0.58 |
| Li | $2s^1$ | 1.08 |
| Na | $3s^1$ | 1.16 |
| Cu | $3d^{10}\ 4s^1$ | 1.22 |
| Mg | $3s^2$ | 1.06 |
| Zn | $3d^{10}\ 4s^2$ | 1.22 |
| F | $2s^2\ 2p^5$ | 0.80 |

## Competing Cu(II) oxides with various magnetic orderings

Potential competing phases containing Cu(II) with the $3d^9$ open-shell configuration have been particularly considered with relevant magnetic configurations in this work. Specifically, they are binary and ternary oxides with previously reported antiferromagnetic (AFM) orderings, as summarised in Table S2. Since the spin-orbit coupling (SOC) effect only incorporated energy differences of the order of a few meV/atom, SOC was included only if these oxides exist as stable phases in the chemical potential diagrams.

**Table S2.** Potential competing binary and ternary Cu(II) oxides with various AFM orderings.

| | Space group | AFM Bravais cell | Magnetisation axis | References |
|---|---|---|---|---|
| CuO | $C2/c$ | $\sqrt{2}a, b, \sqrt{2}c$ | $[10\bar{1}]$ | Calc. [1] |
| CuO | $C2/c$ | $2a, b, 2c$ | [010] | Exp. [2–4] |
| CuO | $C2/c$ | $a, b, 2c$ | [010] | Exp. [5] |
| $Cu_4O_3$ | $I4_1/amd$ | $2a, 2b, c$ | | Calc. [6] |
| $CuTeO_4$ | $P2_1/n$ | $a, b, c$ | [001] | Calc. [7] |
| $Cu_3TeO_6$ | $Ia\bar{3}$ | $a, b, c$ | [001] | Exp. [8], Calc. [9] |

**Band gap analysis**

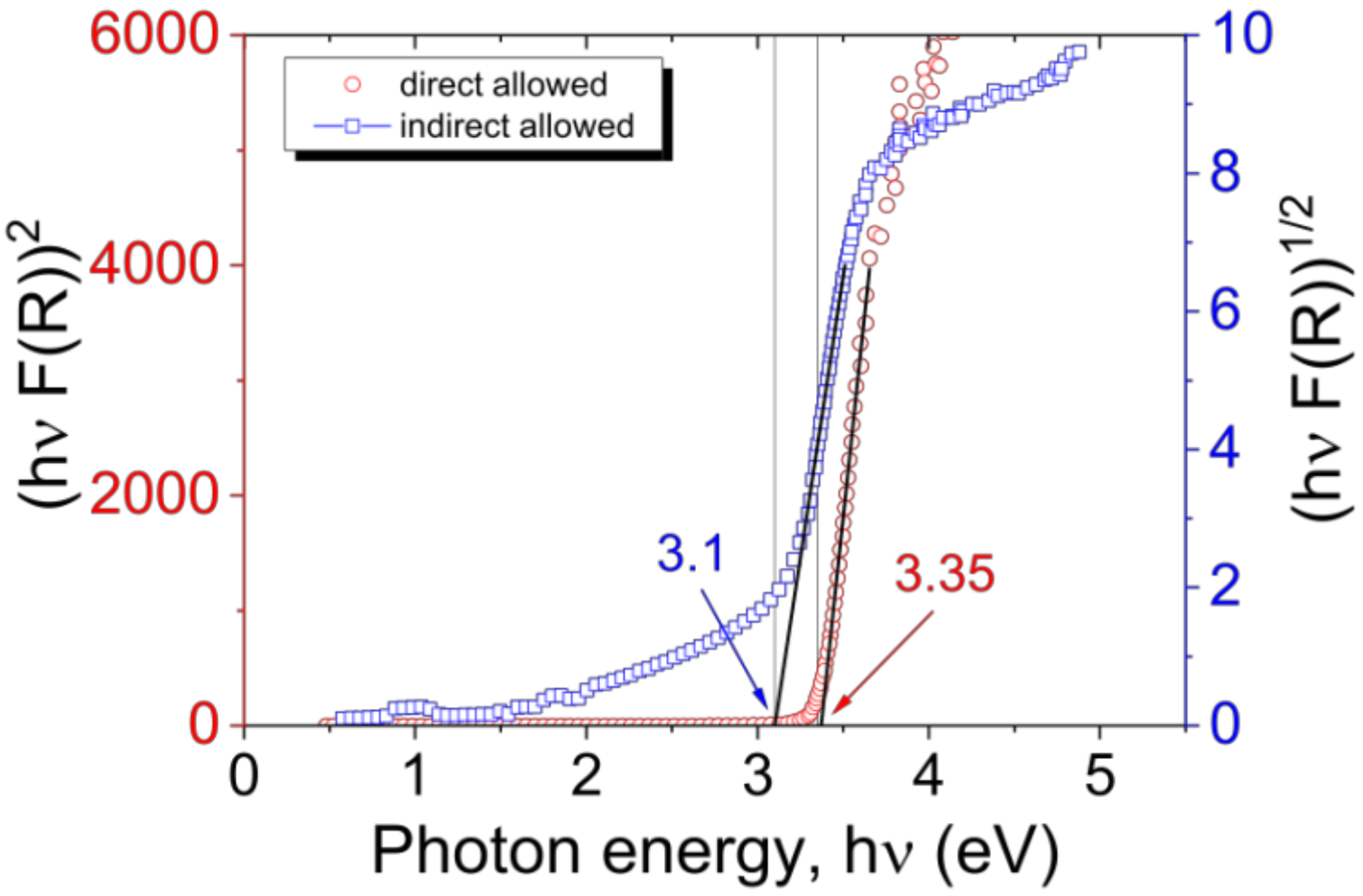

**Fig. S1.** Reanalysis of the diffuse reflectance spectrum reported by Roginskii *et al*. [10]. Direct and indirect band gaps of 3.35 and 3.10 eV, respectively, have been obtained by fitting the corresponding plots.

**Native point defects in $TeO_3$ under O-poor conditions**

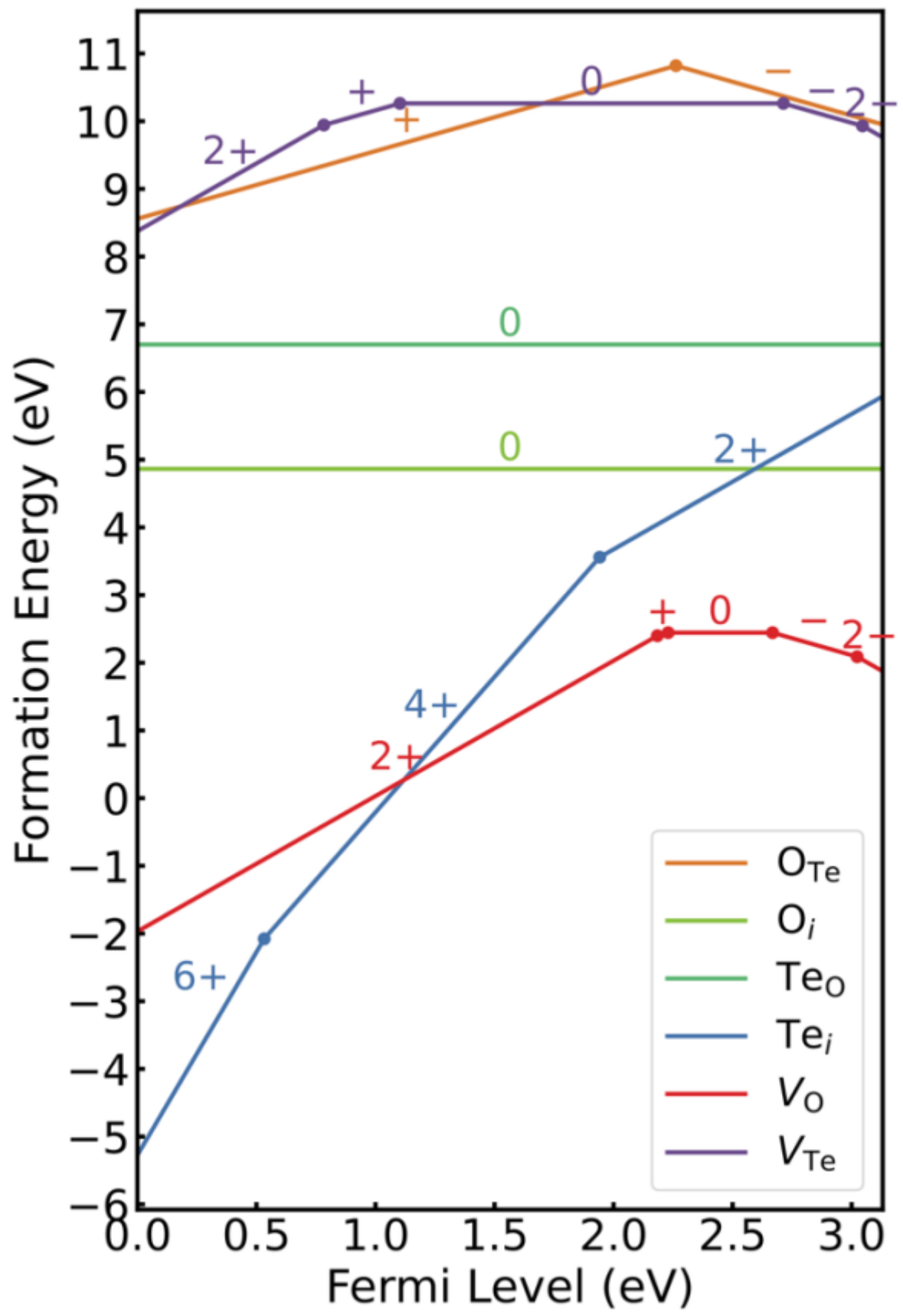

**Fig. S2.** Formation energies of native point defects as a function of the Fermi level under the O-poor conditions. $V_X$ and $X_i$ ($X$ = Te or O) denote vacancies and interstitials, respectively. $X_Y$ are antisites with $X$ taking a $Y$ site ($X$, $Y$ = Te or O). The lower and upper limits of the Fermi level are the calculated VBM and CBM, respectively. For each defect type, only its formation energy in the most stable charge state is shown at a given Fermi level; the slope reflects the charge state. A similar O-vacancy formation energy is observed to that under O-rich conditions.

### *β*-$TeO_3$ supercell for the test of STE formation

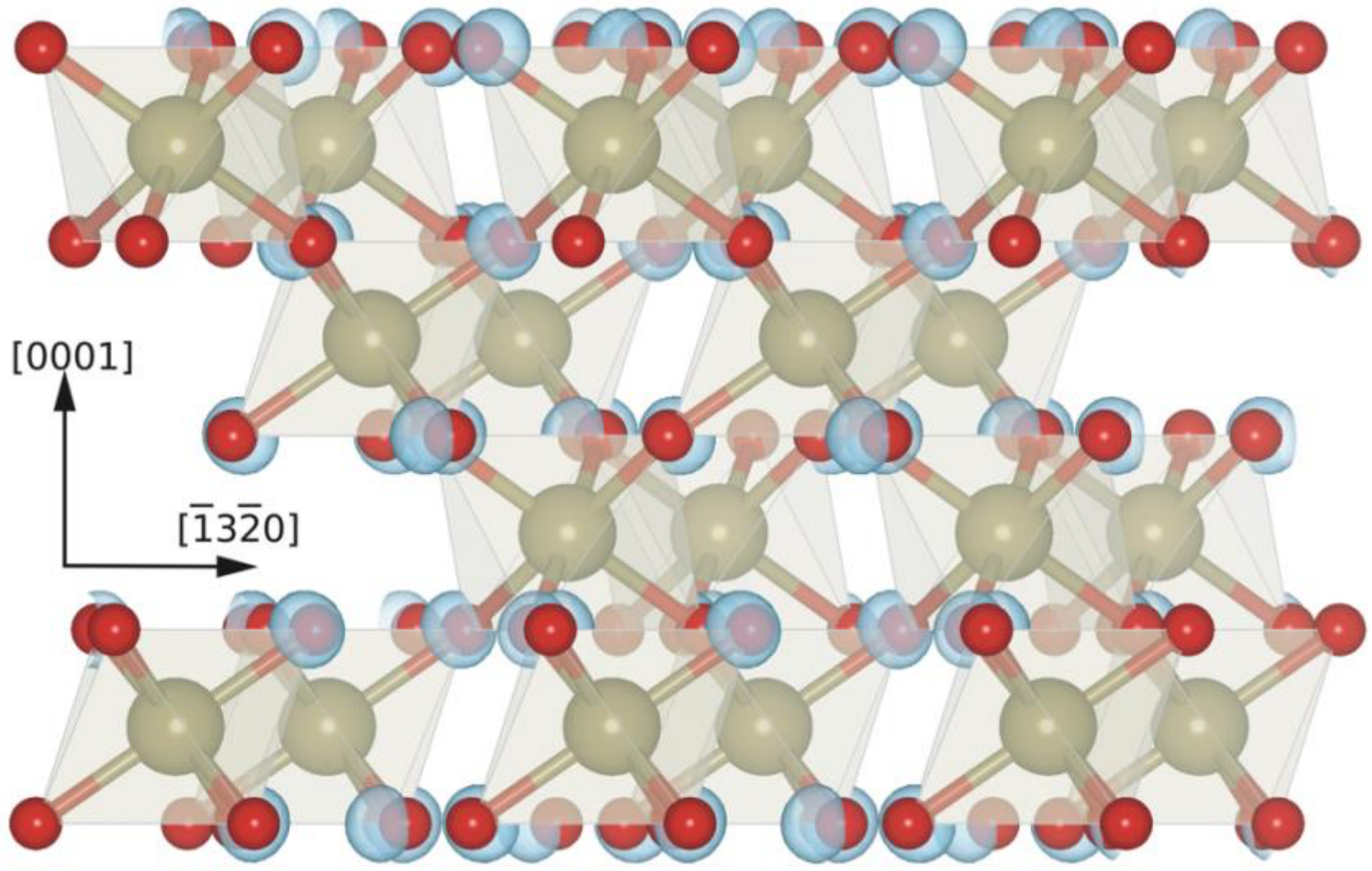

**Fig. S3.** Relaxed atomic configuration of the supercell containing one additional electron to test the formation of an STE, together with the squared wavefunction of the single-particle electronic state of interest located near the CBM, enclosed by isosurfaces of 0.001 1/Å$^3$ coloured in cyan. The Te atoms are in the khaki colour and O in red. No obvious electron localisation is observed.

## Chemical potentials of impurities

**Table S3.** Chemical potentials of $X$ = H, F, Li, Na, Cu, Mg, and Zn obtained from their corresponding solubility limits (competing phase formation) as impurities in $TeO_3$.

| $X$ | Chemical potential conditions | | | |
|---|---|---|---|---|
| | O rich | | O poor | |
| | $\Delta\mu_X$ (eV) | Competing phase | $\Delta\mu_X$ (eV) | Competing phase |
| H | −1.69 | $Te(OH)_6$ | −1.60 | $TeHO_3$ |
| F | −1.89 | $Te_5(O_2F_{11})_2$ | −2.03 | $TeOF_2$ |
| Li | −3.95 | $Li_2TeO_4$ | −3.85 | $Li_2TeO_4$ |
| Na | −3.79 | $Na_2Te_2O_7$ | −3.69 | $Na_2Te_2O_7$ |
| Cu | −2.12 | $CuTeO_4$ | −1.93 | $CuTeO_4$ |
| Mg | −6.71 | $MgTeO_4$ | −6.52 | $MgTeO_4$ |
| Zn | −3.55 | $Zn_3TeO_6$ | −3.36 | $Zn_3TeO_6$ |

**All considered impurity configurations in *β*-$TeO_3$ under O-rich conditions**

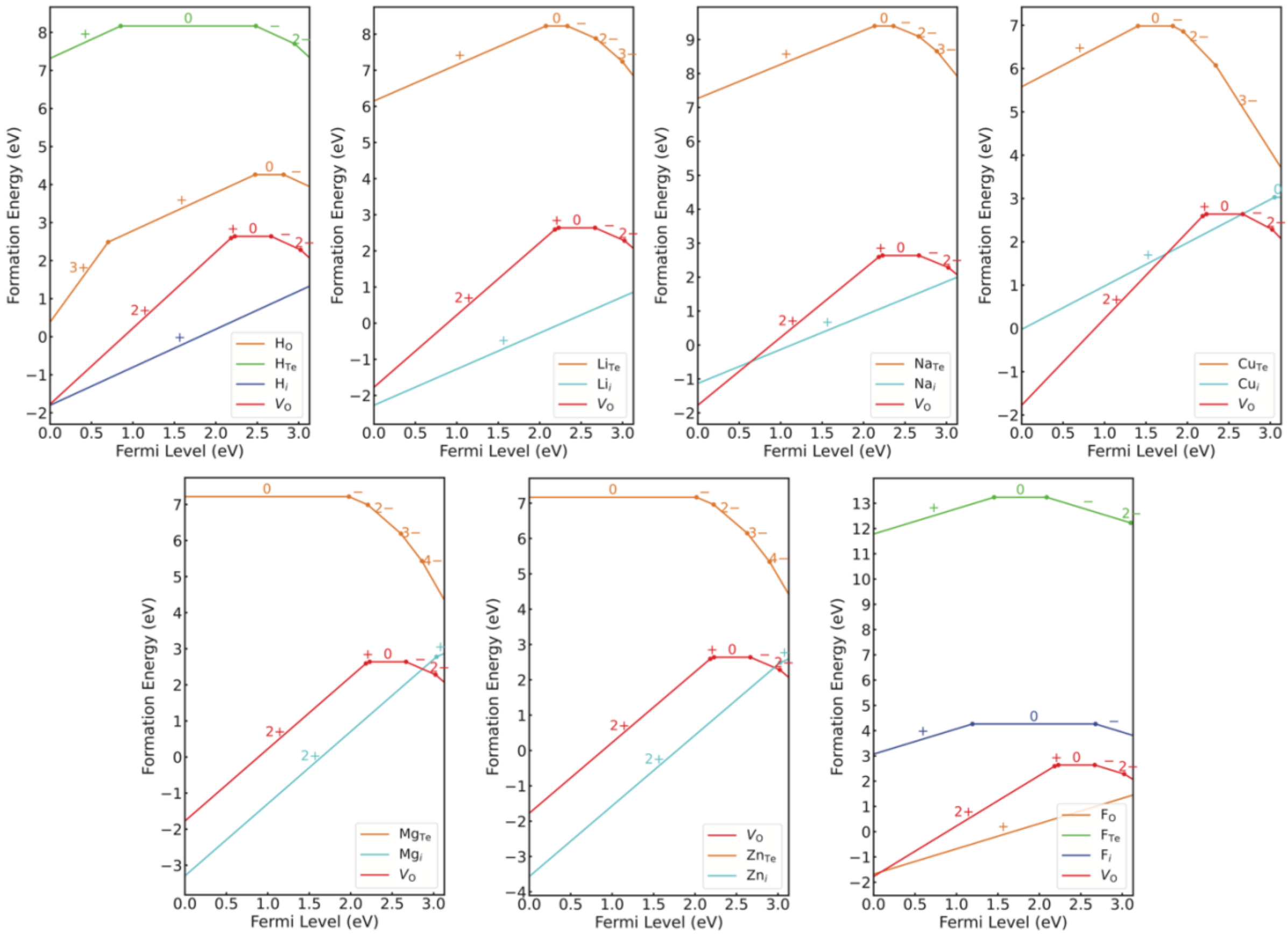

**Fig. S4.** Formation energies of all considered impurity configurations in *β*-$TeO_3$, together with that of the O vacancy, as a function of the Fermi level under O-rich conditions. $X_i$, $X_{Te}$, and $X_O$ ($X$ = H, Li, Na, Cu, Mg, Zn, or F) denote the impurities at interstitial, Te-substitutional, and O-substitutional sites, respectively. Their formation energies and transition levels are shown similarly to those in Fig. S2.

## Impurities in $TeO_3$ under O-poor conditions

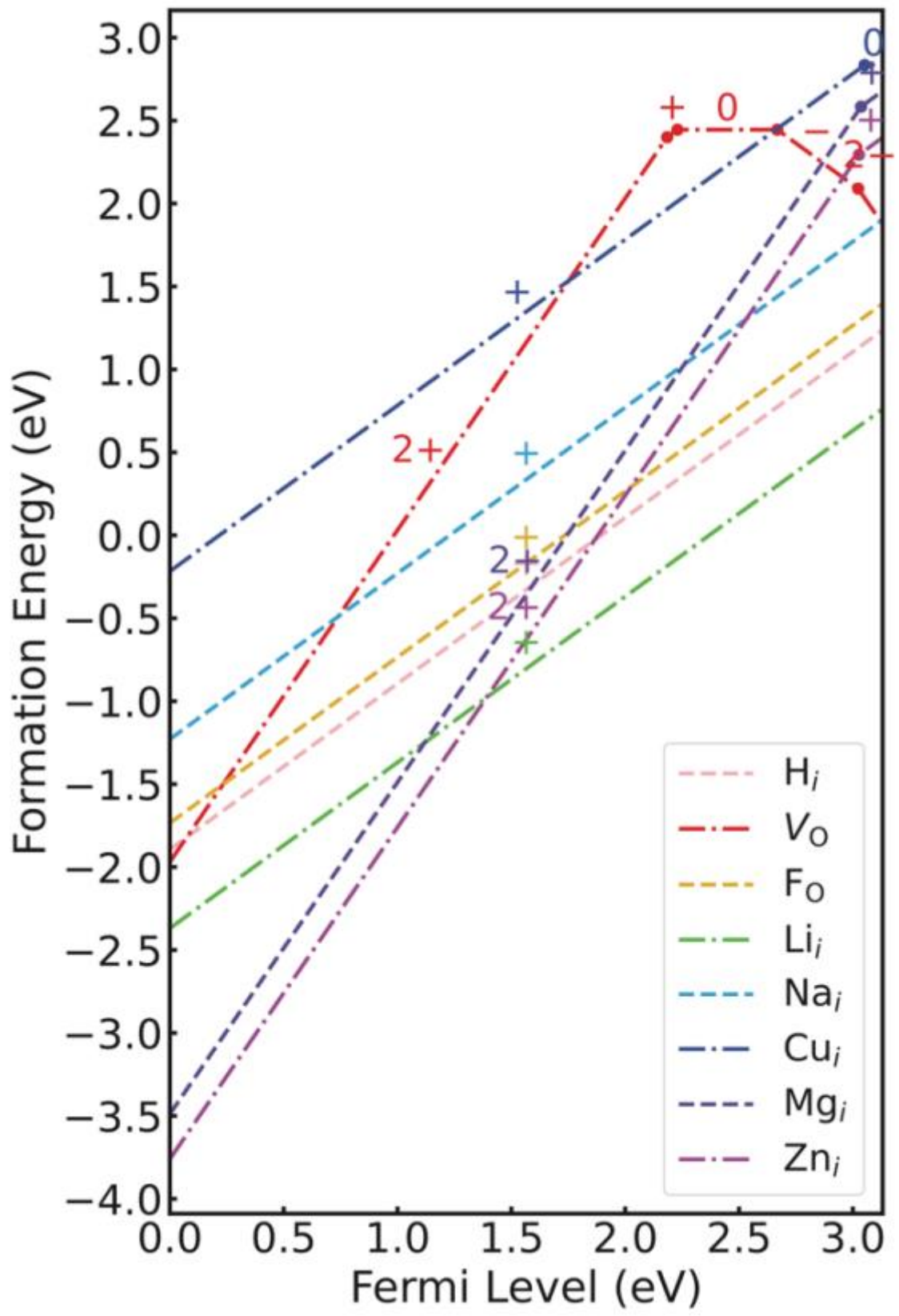

**Fig. S5.** Formation energies of the impurities in $TeO_3$, together with that of the O vacancy, as a function of the Fermi level under O-poor conditions. $X_i$ and $X_O$ ($X$ = H, Li, Na, Cu, Mg, Zn, or F) denote the impurities at interstitial and O sites, respectively. Their formation energies and transition levels are shown similarly to those in Fig. S2.

## Equilibrium carrier concentrations and Fermi levels under O-rich conditions

**Table S4.** Equilibrium carrier/defect concentrations (1/cm$^3$) and Fermi levels (eV) relative to the VBM in $\beta$-$TeO_3$, undoped or doped, at 300 K.

| | Fermi level | $n$ | $p$ | Major defect concentrations |
|---|---|---|---|---|
| Undoped | 1.74 | 1.3E−05 | 4.7E−09 | $V_O^{2+}$ (6.4E−06) |
| H | 2.61 | 5.8E+09 | 1.0E−23 | $H_i^+$ (5.8E+09) |
| Li | 2.84 | 4.2E+13 | 1.4E−27 | $Li_i^+$ (4.2E+13) |
| Na | 2.27 | 1.1E+04 | 5.3E−18 | $Na_i^+$ (1.1E+04) |
| Cu | 1.74 | 1.7E−05 | 3.4E−09 | $Cu_i^+$ (1.0E−05), $V_O^{2+}$ (3.6E−06) |
| Mg | 2.24 | 3.6E+03 | 1.6E−17 | $Mg_i^{2+}$ (1.8E+03) |
| Zn | 2.33 | 1.2E+05 | 4.9E−19 | $Zn_i^{2+}$ (6.0E+04) |
| F | 2.55 | 5.4E+08 | 1.1E−22 | $F_O^+$ (5.4E+08) |

**Table S5.** Equilibrium carrier/defect concentrations (1/cm$^3$) and Fermi levels (eV) relative to the VBM in $\beta$-$TeO_3$, undoped or doped, at 700 K.

| | Fermi level | $n$ | $p$ | Major defect concentrations |
|---|---|---|---|---|
| Undoped | 1.83 | 1.1E+10 | 1.2E+08 | $V_O^{2+}$ (5.3E+09), $V_O^+$ (5.9E+07) |
| H | 2.74 | 3.7E+16 | 3.5E+01 | $H_i^+$ (3.7E+16) |
| Li | 2.96 | 1.5E+18 | 8.7E−01 | $Li_i^+$ (1.5E+18) |
| Na | 2.39 | 1.2E+14 | 1.1E+04 | $Na_i^+$ (1.2E+14) |
| Cu | 1.89 | 3.0E+10 | 4.3E+07 | $Cu_i^+$ (2.8E+10), $V_O^{2+}$ (7.0E+08) |
| Mg | 2.33 | 4.2E+13 | 3.1E+04 | $Mg_i^{2+}$ (2.1E+13) |
| Zn | 2.42 | 1.9E+14 | 6.9E+03 | $Zn_i^{2+}$ (9.3E+13) |
| F | 2.68 | 1.3E+16 | 9.8E+01 | $F_O^+$ (1.3E+16) |

**Table S6.** Equilibrium carrier/defect concentrations (1/cm$^3$) and Fermi levels (eV) relative to the VBM in $\beta$-$TeO_3$, undoped or doped, at 300 K (quenched from 700K).

| | Fermi level | $n$ | $p$ | Major defect concentrations |
|---|---|---|---|---|
| Undoped | 2.43 | 6.8E+06 | 8.7E−21 | $V_O^+$ (8.1E+06), $V_O^-$ (1.2E+06) |
| H | 3.01 | 3.7E+16 | 1.6E−30 | $H_i^+$ (3.7E+16) |
| Li | 3.11 | 1.5E+18 | 3.8E−32 | $Li_i^+$ (1.5E+18) |
| Na | 2.86 | 1.2E+14 | 5.1E−28 | $Na_i^+$ (1.2E+14) |
| Cu | 2.65 | 2.8E+10 | 2.1E−24 | $Cu_i^+$ (2.8E+10), $V_O^-$ (3.5E+08) |
| Mg | 2.84 | 4.2E+13 | 1.4E−27 | $Mg_i^{2+}$ (2.1E+13) |
| Zn | 2.88 | 1.9E+14 | 3.2E−28 | $Zn_i^{2+}$ (9.3E+13), $Zn_i^+$ (2.7E+11) |
| F | 2.99 | 1.3E+16 | 4.5E−30 | $F_O^+$ (1.3E+16) |

**Surface energies for the O chemical potential $\Delta\mu_O > 0$ eV**

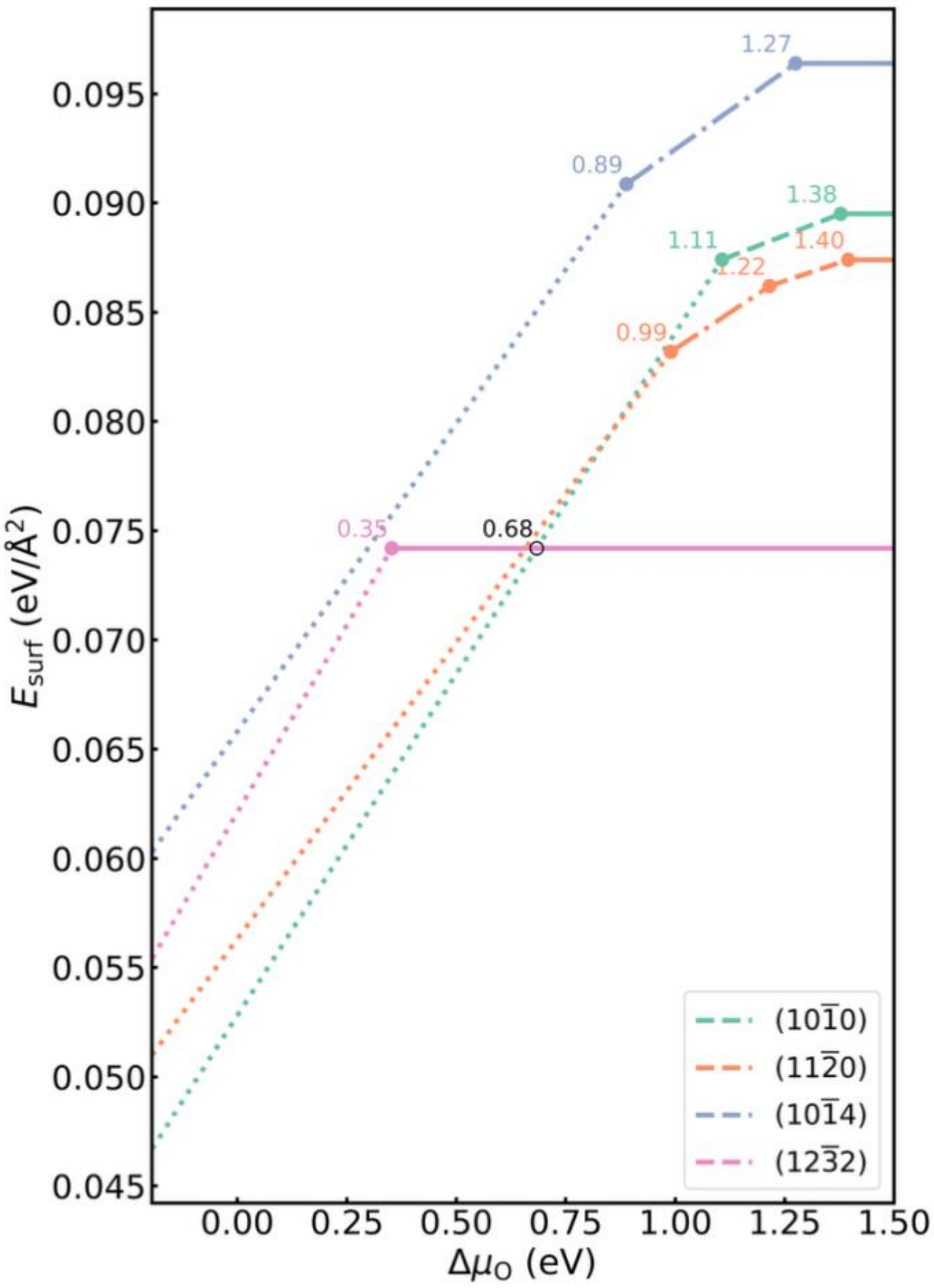

**Fig. S6.** Surface energies of various surface orientation in $\beta$-$TeO_3$ with and without surface O vacancies for $\Delta\mu_O > 0$ eV. The solid, dashed, dash-dotted, and dotted lines represent stoichiometric surfaces, and surfaces with low, medium, and high O-vacancy concentrations, respectively. For each surface orientation, only the surface energy of the most thermodynamically stable O-vacancy configuration is shown at a given O chemical potential. The solid dots represent the transitions, at specific $\Delta\mu_O$ values as printed in the figure, between different O-vacancy configurations of the same surface orientation. The major transition across different surface orientations is marked by a black circle at $\Delta\mu_O = 0.68$ eV.

## References

[1] A. Filippetti, V. Fiorentini, Magnetic Ordering in CuO from First Principles: A Cuprate Antiferromagnet with Fully Three-Dimensional Exchange Interactions, Phys. Rev. Lett. 95 (2005) 086405. https://doi.org/10.1103/PhysRevLett.95.086405.

[2] B.X. Yang, J.M. Tranquada, G. Shirane, Neutron scattering studies of the magnetic structure of cupric oxide, Phys. Rev. B 38 (1988) 174–178. https://doi.org/10.1103/PhysRevB.38.174.

[3] B.X. Yang, T.R. Thurston, J.M. Tranquada, G. Shirane, Magnetic neutron scattering study of single-crystal cupric oxide, Phys. Rev. B 39 (1989) 4343–4349. https://doi.org/10.1103/PhysRevB.39.4343.

[4] J.B. Forsyth, P.J. Brown, B.M. Wanklyn, Magnetism in cupric oxide, J. Phys. C 21 (1988) 2917–2929. https://doi.org/10.1088/0022-3719/21/15/023.

[5] Yu.G. Raydugin, V.E. Naish, E.A. Turov, On the magnetic structure of cupric oxide (CuO), J. Magn. Magn. Mater. 102 (1991) 331–338. https://doi.org/10.1016/0304-8853(91)90147-3.

[6] M.H. Whangbo, H.J. Koo, Spin dimer analysis of the spin exchange interactions in paramelaconite $Cu_4O_3$ and its analogue $Ag_2Cu_2O_3$ and the spin ordering of the $Cu_2O_3$ spin lattice leading to their magnetic phase transitions, Inorg. Chem. 41 (2002) 3570–3577. https://doi.org/10.1021/ic020141x.

[7] A.S. Botana, M.R. Norman, Electronic structure of $CuTeO_4$ and its relationship to cuprates, Phys. Rev. B 95 (2017) 115123. https://doi.org/10.1103/PhysRevB.95.115123.

[8] M. Månsson, K. Prša, J. Sugiyama, D. Andreica, H. Luetkens, H. Berger, Magnetic Order and Transitions in the Spin-web Compound $Cu_3TeO_6$, Phys. Procedia 30 (2012) 142–145. https://doi.org/10.1016/j.phpro.2012.04.059.

[9] J. Chakraborty, Ab initio calculations on three dimensional antiferromagnet $Cu_3TeO_6$, J. Phys. Chem. Solids 134 (2019) 182–186. https://doi.org/10.1016/j.jpcs.2019.05.038.

[10] E.M. Roginskii, V.G. Kuznetsov, M.B. Smirnov, O. Noguera, J.R. Duclère, M. Colas, O. Masson, P. Thomas, Comparative Analysis of the Electronic Structure and Nonlinear Optical Susceptibility of $\alpha$-$TeO_2$ and $\beta$-$TeO_3$ Crystals, J. Phys. Chem. C 121 (2017) 12365–12374. https://doi.org/10.1021/acs.jpcc.7b01819.